\documentclass{cupconf}
\usepackage{graphicx}
\usepackage{natbib}

  \checkfont{eurm10}
  \iffontfound
    \IfFileExists{upmath.sty}
      {\typeout{^^JFound AMS Euler Roman fonts on the system,
                   using the 'upmath' package.^^J}%
       \usepackage{upmath}}
      {\typeout{^^JFound AMS Euler Roman fonts on the system, but you
                   dont seem to have the}%
       \typeout{'upmath' package installed. cupconf.cls can take advantage
                 of these fonts,^^Jif you use 'upmath' package.^^J}%
      }
  \else
  \fi

  \checkfont{msam10}
  \iffontfound
    \IfFileExists{amssymb.sty}
      {\typeout{^^JFound AMS Symbol fonts on the system, using the
                'amssymb' package.^^J}%
       \usepackage{amssymb}%
       \let\le=\leqslant  
       \let\ge=\geqslant  
      }{}
  \fi

  \IfFileExists{amsbsy.sty}
    {\typeout{^^JFound the 'amsbsy' package on the system, using it.^^J}%
     \usepackage{amsbsy}}
    {}

\newsavebox{\astrutbox}
\sbox{\astrutbox}{\rule[-5pt]{0pt}{20pt}}

\def\note #1]{{\bf #1]}}
\def\dd{{\rm d}}
\def\bolddelta{\delta\kern-0.45em\delta\kern-0.45em\delta}
\def\boldr{\mbox{\boldmath$r$}}
\def\bolddelr{\bolddelta \boldr}
\def\muHz{\,\mu{\rm Hz}}
\def\div{{\rm div}\,}
\def\fig{./}
\def\eye{{\rm i}}
\def\boldF{{\bf F}}
\def\CT{{\cal T}}
\def\titnt #1].{\hskip -0.1pt}

\title[Asteroseismology of red giants] 
{Asteroseismology of red giants}

\author[J. Christensen-Dalsgaard]%
{J{\o}rgen Christensen-Dalsgaard$^1$
}

\affiliation{$^1$Department of Physics and Astronomy, Building 1520,
Aarhus University, Ny Munkegade, 8000 Aarhus C, Denmark}

\pubyear{2011}
\volume{538}
\pagerange{1--31}
\date{?? and in revised form ??}
\begin{document}
\maketitle

\begin{abstract}
Red-giant stars are emerging as one of the most interesting areas of
space asteroseismology. 
Even a relatively basic analysis leads to the determination of the global
parameters of the stars, such as their mass and radius, and the very 
extensive space-based data now available for a large number of stars allow
detailed investigation of the deep interiors of red giants, including
distinguishing between stars that do and do not have helium fusion in the
core, on the basis of periods of gravity waves partially trapped in the core.
Here I review the theoretical background for these new developments and provide
a simple explanation for the effect on the period spacing of central
helium fusion.
\end{abstract}

\section{Introduction}

%
\label{sec:intro}
Asteroseismology, and hence the study of stellar properties, 
is being revolutionized by the extremely accurate and extensive data from
the CoRoT \citep{Baglin2009} and {\it Kepler} \citep{Boruck2009} space
missions.
Analysis of timeseries of unprecedented extent, continuity and sensitivity
have allowed the study of a broad range of stellar variability,
including oscillations of a variety of pulsating stars
\citep[see, e.g.,][for reviews]{Gillil2010, Christ2011},
and leading to comparative asteroseismology (or {\it synasteroseismology}%
\footnote{a term coined by D. O. Gough}%
)
for main-sequence stars showing solar-like oscillation \citep{Chapli2011}.
Also, early analyses of {\it Kepler} data have demonstrated 
the power of asteroseismology in characterizing the central stars in
planetary systems \citep{Christ2010, Batalh2011}, 
and such investigations will undoubtedly play a major role in the 
continuing {\it Kepler} exploration of extra-solar planetary systems.

However, perhaps the most striking results of space asteroseismology have
come from the investigation of red giants.
Given the extensive outer convection zones of red giants, solar-like
oscillations were predicted quite early \citep{Christ1983}.
Ground-based observations have been carried out in a few cases
\citep[e.g.,][]{Frands2002, DeRidd2006} but owing to the very long periods of
these huge stars such observations are extremely demanding in terms of
observing and observer's time.
Space observations, on the other hand, allow nearly continuous observations
over very extended periods, as demonstrated by early observations by the
WIRE \citep{Retter2003} and MOST \citep{Barban2007} satellites.
With CoRoT and {\it Kepler} we have obtained extensive data for literally
thousands of red giants, providing precise characterization of their
overall characteristics, as well as, 
in some cases, detailed information on the properties of their deep interiors.
This promises completely to change our understanding of these late phases
of stellar evolution.

Here I provide an overview of the oscillation properties of red giants 
and discuss some of the initial results of the analysis of data from
CoRoT and {\it Kepler}.
An introduction to the properties of solar-like oscillations,
emphasizing the observational aspects, was provided by Bedding (this volume).
Further introductions to solar-like stellar oscillations were provided
by \citet{Christ2004} and, in particular, by the monograph by
\citet{Aerts2010}.
To illustrate the relevant properties, computations of red-giant models 
and oscillations are presented.
These probably do not fully cover the complexities of these late stages
of stellar evolution and hence are not intended as models of specific
observations, but it is hoped that they at least capture the main features
of these stars.
The modelling was carried out using the Aarhus STellar Evolution Code ASTEC
\citep{Christ2008a},
with OPAL equation of state \citep{Rogers1996} and opacities \citep{Iglesi1996},
and NACRE \citep{Angulo1999} nuclear reaction rates.
Convection was treated with the \citet{Bohm1958} mixing-length formalism,
with a mixing length proportional to the pressure scale height and
approximately calibrated to a solar model.
Diffusion and settling were neglected and no mass loss was included.
Adiabatic oscillations were calculation using the Aarhus adiabatic pulsation
code ADIPLS \citep{Christ2008b}.
To resolve the very rapid variations in the eigenfunctions in the stellar
core, the models were transferred from the original mesh used in the model
calculation to a much finer mesh, designed to capture adequately
the properties of the eigenfunctions.
However, it is probably fair to say that the mesh distribution, and more
generally the numerical procedures to deal with this type of oscillation,
requires further optimization and testing.


\section{Simple properties of stellar oscillations}

%
\label{sec:properties}
This section provide a brief background on the theory of stellar oscillations,
as far as the oscillation frequencies and their relation to stellar 
properties are concerned.
Some emphasis is given to the oscillation properties of red giants.

To understand the properties of stellar oscillation frequencies, it is a 
good approximation to assume that the oscillations are adiabatic, and
hence ignore the processes that damp or excite the modes.
This approximation breaks down in the near-surface layers, where 
there are strong departures from adiabaticity.
Also, the modelling of this region is highly uncertain, owing to the detailed
effects of convection on the mean structure of the star and the dynamics 
of the oscillations.
These near-surface processes have a significant effect on the frequencies,
as is evident in the analysis of solar observations, and they must be taken
into account in detailed asteroseismic analyses of solar-like oscillation
frequencies \citep[e.g.,][]{Kjelds2008}.
However, for the present discussion we can ignore them.

With this approximation the problem of computing stellar 
oscillation frequencies for a given stellar model is 
relatively straightforward.%
\footnote{Although, as discussed below some care is
needed when considering red giants.}
This feature, together with the ability to determine the 
observed frequencies with very high accuracy, makes stellar oscillations
such excellent diagnostics of the stellar interior.
For the purpose of computing adiabatic frequencies, and assuming that
the equilibrium model satisfies hydrostatic equilibrium
(see Kawaler, this volume),
the structure of the star is characterized by the density $\rho(r)$ as
a function of distance $r$ to the centre, as well as the adiabatic exponent
$\Gamma_1 (r)$, where
\begin{equation}
\Gamma_1 = \left( {\partial \ln p \over \partial \ln \rho} \right)_{\rm ad} 
\; ,
\end{equation}
where $p$ is pressure and the derivative corresponds to an adiabatic change.
Given $\rho(r)$ the mass distribution in the star can immediately 
be determined,
and $p(r)$ can then be computed from the equation of hydrostatic equilibrium,
with a suitable boundary condition.
This essentially defines all that is needed to complete the equations of
adiabatic oscillation.
For the predominantly acoustic solar-like oscillations the most important
quantity is the adiabatic sound speed $c$, given by
\begin{equation}
c^2 = {\Gamma_1 p \over \rho} \; .
\label{eq:csqfull}
\end{equation}
In many relevant cases, the equation of state can be approximated by
the ideal gas law:
\begin{equation}
p = {k_{\rm B} \rho T \over \mu m_{\rm u}} \; ,
\label{eq:idealg}
\end{equation}
where $T$ is temperature, $\mu$ the mean molecular weight,
$k_{\rm B}$ Boltzmann's constant and $m_{\rm u}$ the atomic mass unit.
We then obtain
\begin{equation}
c^2 \simeq {\Gamma_1 k_{\rm B} T \over \mu m_{\rm u}} \; .
\label{eq:csqig}
\end{equation}

We consider small-amplitude oscillations in a slowly rotating star.
Then the dependence of the eigenfunctions on co-latitude $\theta$ and
longitude $\phi$ can be separated as spherical harmonics 
$Y_l^m(\theta, \phi)$;
the degree $l$ measures the total number of nodal lines on the stellar
surface and provides a measure of the local horizontal wavenumber 
$k_{\rm h}$:
\begin{equation}
k_{\rm h}^2 = {l(l+1) \over r^2} \; ,
\end{equation}
while the azimuthal order $m$, with $|m| \le l$,
measures the number of nodal lines crossing the equator.
For each $(l, m)$ the star has a set of eigenfrequencies $\omega_{nlm}$ 
labelled by the radial order $n$.
For adiabatic oscillations $\omega_{nlm}$ is real;
the oscillation depends on time $t$ and $\phi$ as $\cos(m \phi - \omega t)$,
i.e., for $m \neq 0$, as a running wave in longitude.
For a nonrotating star the frequencies are independent of $m$. 
Slow rotation with angular velocity $\Omega$ introduces a frequency splitting
\begin{equation}
\omega_{nlm} = \omega_{nl0} + m \beta_{nlm} \langle \Omega \rangle_{nlm} \; ,
\end{equation}
where $\langle \Omega \rangle_{nlm}$ is an average over the star 
with a weight determined by the oscillation eigenfunction.
If $\omega = \Omega(r)$ depends on $r$ alone $\beta_{nlm} = \beta_{nl}$
and the average are independent of $m$ and the rotational splitting is linear
in $m$.

After separation of $Y_l^m$ the equations of adiabatic oscillations reduce
to a set of four linear differential equations in $r$, which suitable 
boundary conditions at the centre and surface, which can be solved numerically,
with the frequencies $\omega_{nlm}$ as eigenvalues.
However,
a great deal of insight into the properties of the solution and the relation
of the frequencies to the structure of the star can be obtained from
asymptotic analysis of the equations.
\citet{Deubne1984} derived an asymptotic equation, applicable for
modes of high radial order, in terms of the quantity 
$X = c^2 \rho^{1/2} \div \bolddelr$, where $\bolddelr$
is the displacement vector:
\begin{equation}
{\dd^2 X \over \dd r^2} = -K(r) X \; ,
\label{eq:asymp}
\end{equation}
where
\begin{equation}
K = {1 \over c^2} \left[ S_l^2 \left({N^2 \over \omega^2} - 1\right) 
+ \omega^2 - \omega_{\rm c}^2 \right] \; .
\label{eq:kasymp}
\end{equation}
Thus $K$ is controlled by three characteristic frequencies of the star:
the {\it Lamb frequency} $S_l$, with
\begin{equation}
S_l^2 = {l(l+1) c^2 \over r^2} \; ,
\label{eq:lamb}
\end{equation}
the {\it buoyancy frequency} (or {\it Brunt-V\"ais\"al\"a frequency}) $N$,
\begin{equation}
N^2 = g \left( {1 \over \Gamma_1} {\dd \ln p \over \dd r} 
- {\dd \ln \rho \over \dd r} \right) \; ,
\label{eq:buoy}
\end{equation}
where $g$ is the local gravitational acceleration,
and the {\it acoustic cut-off frequency} $\omega_{\rm c}$,
\begin{equation}
\omega_{\rm c}^2 = {c^2 \over 4 H^2} \left(1 - 2 {\dd H \over \dd r} \right)
 \; ,
\end{equation}
where $H = - (\dd \ln \rho / \dd r)^{-1}$ is the density scale height.

Equation (\ref{eq:asymp}) determines the local properties of the
eigenfunction.
In regions where $K(r) > 0$ the solution locally oscillates as a function
of $r$, while the behaviour is exponentially increasing or decreasing where
$K(r) < 0$. 
The intermediate points, where $K(r) = 0$, are called {\it turning points}.
Often there is an interval $[r_1, r_2]$ where $K(r) > 0$, with $K$ being 
negative just outside and the solution decreasing exponentially
in the direction away from that interval.
In that case the mode is said to be trapped between $r_1$ and $r_2$;
from JWKB analysis \citep[e.g.,][]{Gough2007} one finds that 
the frequency approximately satisfies the dispersion relation
\begin{equation}
\int_{r_1}^{r_2} K^{1/2} \dd r = (n - 1/2) \pi \; .
\label{eq:jwkb}
\end{equation}

The behaviour of the Lamb frequency and the acoustic cut-off frequency
is relatively simple. 
The Lamb frequency generally decreases with increasing $r$;%
\footnote{possibly with the exception of discontinuities in composition,
caused by nuclear burning in the stellar core, which leads to discontinuities in
density and hence sound speed}
for the low degrees that are relevant to stellar observations it is small 
near the surface. 
Here, typically, $\omega_{\rm c}^2$ dominates in Eq.~(\ref{eq:kasymp}).
In the atmosphere, where the temperature is approximately constant, 
$H$ can be approximated by the (constant) pressure scale height and we obtain
\begin{equation}
\omega_{\rm c} \simeq \omega_{\rm c, atm} 
= {1 \over 2} g \sqrt{ \Gamma_1 \rho / p}
\propto M R^{-2} T_{\rm eff}^{-1/2} \; ,
\label{eq:cutoff}
\end{equation}
evaluated in the atmosphere;
in the last proportionality we assumed the ideal gas law 
(see Eq.~\ref{eq:idealg}) and that
the atmospheric temperature is proportional to the effective temperature
$T_{\rm eff}$.
Below the surface $\omega_{\rm c}$ decreases rapidly with increasing depth.
When $\omega < \omega_{\rm c, atm}$ the solution decreases with height
in the atmosphere; 
this corresponds to reflection of the mode at the surface so that it is
trapped in the stellar interior.
Waves with $\omega > \omega_{\rm c, atm}$ can propagate out through 
the atmosphere and hence lose energy, leading to strong damping.
Thus $\omega_{\rm c, atm}$ defines the upper limit in frequency to trapped
modes, at least in the adiabatic approximation.

The behaviour of the buoyancy frequency is more complex. 
This is seen most clearly by assuming the ideal gas approximation,
Eq.~(\ref{eq:idealg}).
Then we obtain, from Eq.~(\ref{eq:buoy}), that
\begin{equation}
N^2 = g^2 {\rho \over p} (\nabla_{\rm ad} - \nabla + \nabla_\mu ) \; ,
\label{eq:abuoy}
\end{equation}
where, following the usual convention
(see Kawaler, this volume),
\begin{equation}
\nabla = {\dd \ln T \over \dd \ln p} \; , \qquad
\nabla_{\rm ad} = \left({\partial \ln T \over \partial \ln p} \right)_{\rm ad}
\; , \qquad
\nabla_\mu = {\dd \ln \mu \over \dd \ln p} \; .
\label{eq:grads}
\end{equation}
Since typically $\mu$ increases towards the centre
and hence with increasing $p$, as a result of nuclear burning, 
the term in $\nabla_\mu$ typically gives a positive contribution to
$N^2$ in the deep interior of the star.
It should also be noted that the condition for convective instability%
\footnote{strictly speaking the Ledoux criterion}
is that $N^2 < 0$.
In convection zones the composition is uniform and $\nabla_\mu = 0$;
also, except near the surface $\nabla$ is nearly adiabatic and $N^2$
is only slightly negative.

\begin{figure}
\begin{center}
\includegraphics[width=8.0cm]{\fig/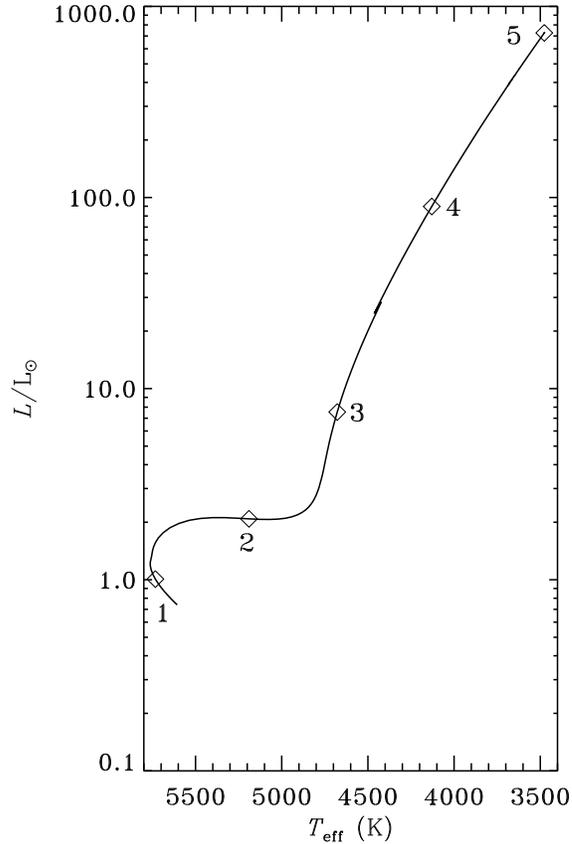}
\caption{Evolution track for a $1 M_\odot$ model, approximately
calibrated to the Sun at the present solar age.
}
   \label{fig:hr1}
\end{center}
\end{figure}

\begin{figure}
\begin{center}
\includegraphics[width=10.0cm]{\fig/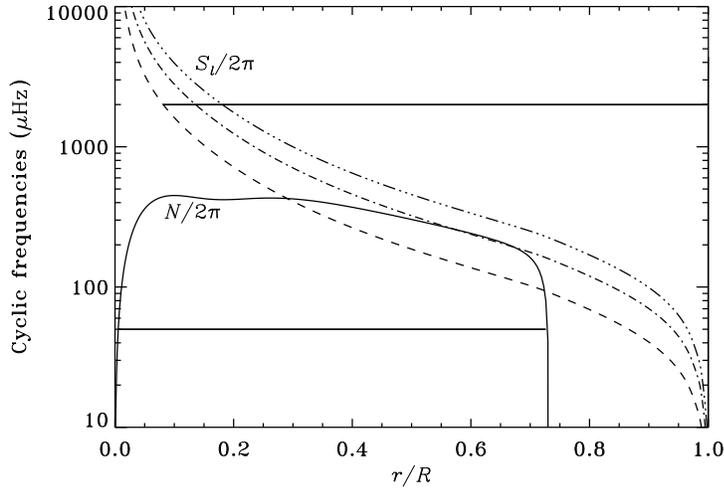}
\caption{Characteristic cyclic frequencies for Model~1 in
the $1 M_\odot$ evolution sequence illustrated in Fig.~\ref{fig:hr1}.
The solid curve shows $N/2 \pi$, where $N$ is the buoyancy frequency
(cf.\ Eq.~\ref{eq:buoy}), whereas the dashed, dot-dashed and triple-dot-dashed
curves show the Lamb frequency $S_l/2 \pi$ (cf.\ Eq.~\ref{eq:lamb})
for $l = 1$, 2 and 3, respectively.
The horizontal solid lines show typical propagation regions for a
p mode (with $l = 1$, at high frequency) and a g mode (at low frequency).
}
   \label{fig:charfr1}
\end{center}
\end{figure}

Solar-like oscillations are believed to be excited stochastically
by the vigorous motion in the near-surface layers of convective envelopes
(see Section~\ref{sec:excitation} below).
Thus stars showing such oscillations have outer convection zones
and hence the buoyancy frequency is positive only in the deeper interior.
Here we illustrate the characteristic frequencies and their dependence on
the evolutionary state by considering selected models in a $1\,M_\odot$
evolution sequence, illustrated in an HR diagram in Fig.~\ref{fig:hr1}.
A typical example of a rather unevolved star, at roughly
solar age, is shown in Fig.~\ref{fig:charfr1}.
Here there is a fairly clear separation between $S_l$ and $N$.
As a result there are essentially two different scenarios for having 
a positive $K$, also illustrated in the figure: 
by having $\omega > S_l$ (and hence typically $\omega \gg N$)
or by having $\omega < N$ (and hence typically $\omega \ll S_l$).
The former case corresponds to acoustic modes (or p modes), where
the restoring force is predominantly pressure,
while the latter case corresponds to internal gravity waves (or g modes),
where the restoring force is buoyancy.

In discussing the properties of these modes in more detail, we
neglect $\omega_{\rm c}$ in Eq.~(\ref{eq:kasymp}), except insofar as to use
that it causes the required reflection at the surface of the p modes.
For the p modes, assuming that $\omega \gg N$, we then obtain from
Eq.~(\ref{eq:kasymp}) that
\begin{equation}
K \simeq {1 \over c^2} (\omega^2 - S_l^2) \; ,
\label{eq:pkasymp}
\end{equation}
with a lower turning point $r_{\rm t}$ defined by $S_l = \omega$ or
\begin{equation}
{c(r_{\rm t}) \over r_{\rm t}} = {\omega \over L} \; ,
\label{eq:rt}
\end{equation}
where $L^2 = l(l+1)$.
Also, the corresponding approximation to Eq.~(\ref{eq:jwkb}) can be
rearranged to yield
\begin{equation}
\int_{r_{\rm t}}^R \left(1 - {L^2 c^2 \over r^2 \omega^2} \right)^{1/2}
{\dd r \over c} = {(n + \alpha) \pi \over \omega} \; .
\label{eq:duvall}
\end{equation}
Here we replaced the $- 1/2$ in Eq.~(\ref{eq:jwkb}) by $\alpha$,
which generally is a function of frequency,
to account for the phase change in the reflection at the surface.
Equation~(\ref{eq:duvall}) has played a major role in the analysis
of solar oscillation frequencies;
a relation of this form was first obtained by \citet{Duvall1982} from
analysis of observed frequencies.
It explicitly demonstrates that in this limit the frequencies are
determined by the sound speed and by $\alpha$.

In unevolved stars $r_{\rm t}$ is close to the centre where $c$
varies relatively slowly.
Equation~(\ref{eq:duvall}) can then be expanded \citep{Gough1986, Gough1993}
to yield, in terms of the cyclic frequencies $\nu_{nl} = \omega_{nl}/2 \pi$,
\begin{equation}
\nu_{nl} = \Delta \nu \left(n + {l \over 2} + \alpha + {1 \over 4} \right) 
- d_{nl} \; ,
\label{eq:tass}
\end{equation}
where 
\begin{equation}
\Delta \nu = \left(2 \int_0^R {\dd r \over c} \right)^{-1} \; ,
\label{eq:largesep}
\end{equation}
and we replaced $L$ by $l + 1/2$ 
\citep[see also][]{Vandak1967, Tassou1980, Tassou1990}.
Neglecting the small correction term $d_{nl}$ this shows that the
frequencies are uniformly spaced in radial order $n$, 
with a spacing given by the {\it large separation} $\Delta \nu$ and
such that there is degeneracy between $\nu_{nl}$ and $\nu_{n-1\,l+2}$.
This degeneracy is lifted by $d_{nl}$ which asymptotically,
for main-sequence models, can be approximated by 
\begin{equation}
d_{nl} \simeq - {\Delta \nu \over 4 \pi^2 \nu_{nl}}L^2
\int_0^R {\dd c \over \dd r} {\dd r \over r} \; .
\label{eq:smallsep}
\end{equation}
To this approximation $d_{nl}$, and hence the small separations
$\delta \nu_{l\,l+2}(n) = \nu_{nl} - \nu_{n-1\,l+2}$, are very sensitive to the
sound-speed gradient in the stellar core and hence,
according to Eq.~(\ref{eq:csqig}),
to the composition profile that has resulted from the nuclear burning.
Thus for main-sequence stars the small frequency separations provide
a measure of the evolutionary stage and hence the age of the star.
Interestingly, although the derivation sketched above and the expression
for $d_{nl}$ do not hold for highly evolved stars such as red giants,
the frequency pattern reflected in Eq.~(\ref{eq:tass}) is still found
(see also Section~\ref{sec:rgastero}).

In the opposite extreme, where $\omega \ll S_l$ and $\omega < N$,
we approximate $K$ by 
\begin{equation}
K \simeq {L^2 \over r^2} \left( {N^2 \over \omega^2} - 1 \right) \; .
\label{eq:gkasymp}
\end{equation}
With a relatively simple behaviour of $N$ such as illustrated in 
Fig.~\ref{fig:charfr1} there are typically just two turning points
$r_1, r_2$ where $\omega = N$,
and Eq.~(\ref{eq:jwkb}) yields
\begin{equation}
L \int_{r_1}^{r_2} \left( {N^2 \over \omega^2} - 1 \right)^{1/2}
{\dd r \over r}  = (n - 1/2) \pi \; .
\label{eq:gfreq}
\end{equation}
If $\omega \ll N$ almost everywhere on $[r_1, r_2]$ we can approximate
Eq.~(\ref{eq:gfreq}) further by neglecting $1$ compared with
$N^2/\omega^2$, to obtain finally an approximate expression for the period
$\Pi = 2 \pi / \omega$:
\begin{equation}
\Pi = {\Pi_0 \over L} (n + \alpha_{\rm g}) \; ,
\label{eq:gper}
\end{equation}
where
\begin{equation}
\Pi_0 = 2 \pi^2 \left(\int_{r_1}^{r_2} N {\dd r \over r}\right)^{-1} \; ,
\label{eq:gpersp}
\end{equation}
assuming that $N^2 \ge 0$ on $[r_1, r_2]$, with $N = 0$ at $r_1$ and
$r_2$
\citep[e.g.,][]{Tassou1980}.
Also, $\alpha_{\rm g}$, which may depend on $l$, reflects the actual
phase shift at the turning points.
Equation~(\ref{eq:gper}) shows that in this case we get periods that are
uniformly spaced in the radial order $n$, with a spacing that is
inversely proportional to $L$.

\begin{figure}
\begin{center}
\includegraphics[width=10.0cm]{\fig/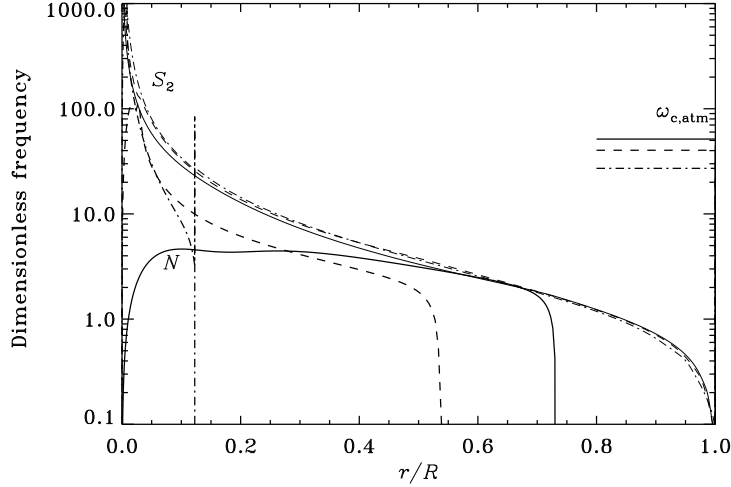}
\caption{Dimensionless characteristic frequencies (in units of
$(G M/R^3)^{1/2}$) for models along the $1 M_\odot$ evolutionary track
illustrated in Fig.~\ref{fig:hr1}.
Solid, dashed and dot-dashed curves are for Models 1, 2 and 3, respectively.
The thick curves show the buoyancy frequency $N$ while the thinner curves
show the Lamb frequency $S_l$, for $l = 2$.
The horizontal lines at the right edge of the plot similarly
show the dimensionless
value of the acoustic cut-off frequency (cf.\ Eq.~\ref{eq:cutoff}) in the
atmosphere of the models.
}
   \label{fig:charfr2}
\end{center}
\end{figure}

The dependence of the characteristic frequencies and hence the oscillation
properties on stellar parameters can to a large extent be characterized
by simple scaling relations.
An important example
is the scaling of the acoustic cut-off frequency in Eq.~(\ref{eq:cutoff}).
Under many circumstances the structure of a star, as a function of
the fractional radius $x = r/R$, approximately satisfies so-called
{\it homology scaling relations} \citep[e.g.,][]{Kippen1990}.%
\footnote{A simple example, where the scalings are exact, are 
polytropic models.}
According to these pressure scales as $G M^2 / R^4$, $G$ being the
gravitational constant, and density scales as $M/R^3$.
It follows that the squared sound speed $c^2$ scales as $G M/R$ and
the squared characteristic frequencies scale as $G M/R^3$.
This motivates introducing the dimensionless scaled frequencies by
\begin{equation}
\hat S_l^2 = {R^3 \over G M} S_l^2 \; , \qquad
\hat N^2 = {R^3 \over G M} N^2 \; , \qquad
\hat \omega_{\rm c}^2 = {R^3 \over G M} \omega_{\rm c}^2 \; ,
\end{equation}
as well as $\hat c^2 = (R/GM) c^2$.
Then the asymptotic equations (\ref{eq:asymp}) and (\ref{eq:kasymp})
can be expressed as
\begin{equation}
{\dd^2 X \over \dd x^2} = - \hat K(r) X \; ,
\label{eq:sasymp}
\end{equation}
where
\begin{equation}
\hat K = {1 \over \hat c^2} \left[ \hat S_l^2 \left({\hat N^2 \over \sigma^2} 
- 1\right) 
+ \sigma^2 - \hat \omega_{\rm c}^2 \right] \; ,
\label{eq:skasymp}
\end{equation}
in terms of the dimensionless frequency
\begin{equation}
\sigma^2 = {R^3 \over G M} \omega^2 \; .
\end{equation}
With similar scalings the full equations of adiabatic oscillations can
be expressed in a form that is homologously invariant.
It follows that homologous models have the same dimensionless frequencies
$\sigma_{nl}$, and that the actual frequencies satisfy the scaling
\begin{equation}
\omega_{nl} \propto (M/R^3)^{1/2} \propto \langle \rho \rangle^{1/2} \; ,
\label{eq:nuscale}
\end{equation}
where $\langle \rho \rangle$ is the mean density of the star.
The same scaling obviously applies, e.g., to the large frequency
separation $\Delta \nu$.

Actual stellar models are never strictly homologous, and in particular
the changing structure of the core as nuclear burning proceeds gives 
strong departures from these scalings.
On the other hand, they are approximately satisfied in the outer
layers which predominantly determine the frequencies of acoustic modes,
and hence for such modes the scaling in Eq.~(\ref{eq:nuscale}) is a
reasonable approximation.
To illustrate this we consider the evolution of the frequencies
along a $1\,M_\odot$ evolution sequence, illustrated in Fig.~\ref{fig:hr1}.
The dimensionless frequencies for Models 1, 2 and 3 are shown in
Fig.~\ref{fig:charfr2}.
Evidently there is relatively little change in $\hat S_2$.
On the other hand, the changes in $\hat N$ are dramatic, showing very 
strong departures from the homologous scaling.
The increase in the depth of the outer convection zone, where $|N| \simeq 0$,
is evident.
Also, $N$ increases strongly in the core.
This is mainly the result of the increase in the gravitational acceleration
(cf.\ Eq.~\ref{eq:abuoy}) as the star develops a very compact
helium core.
Thus Model 3 has a helium core with a mass of $0.18\, M_\odot$,
contained within the inner 0.7 \% of the stellar radius.
In this model the convective envelope is so deep that it extends into a
region where the composition has been changed by hydrogen fusion.
This causes a discontinuity in the composition and hence the density,
giving rise to a spike in the buoyancy frequency, as shown in the figure.

In Fig.~\ref{fig:charfr2} are also indicated the dimensionless 
values $\hat \omega_{\rm c, atm}$ of the acoustic cut-off frequency in
the atmospheres of the three models. 
Since this is determined by the atmospheric properties of the star
it does not follow the homologous frequency scaling.
In fact, using Eq.~(\ref{eq:cutoff})
for $\nu_{\rm c, atm} = \omega_{\rm c, atm}/2 \pi$
and the scaling in Eq.~(\ref{eq:nuscale}) for $\Delta \nu$ it follows that
\begin{equation}
{\nu_{\rm c, atm} \over \Delta \nu} \propto M^{1/2} R^{-1/2} T_{\rm eff}^{-1/2} \; .
\end{equation}
Since $\nu_{\rm c, atm}$ is an upper limit to the frequencies of
modes trapped in the stellar interior, this ratio, according to 
Eq.~(\ref{eq:tass}), provides a measure of the number of acoustic modes
that are trapped.
This obviously decreases as the star evolves up the red-giant branch.

Figure~\ref{fig:charfr2} also shows that in Models 2 and 3 the maximum of
$\hat N$ in the core is substantially higher than $\hat \omega_{\rm c, atm}$.
This means that {\it all} trapped nonradial modes satisfy
$\omega < N, S_l$ in the core and hence behave like g modes there, while
they may behave as acoustic modes in the outer parts of the star where
$\omega > N, S_l$.
The overall character of the mode then depends on whether it has the largest
amplitude in the g-mode or the p-mode region, and hence, effectively,
on the behaviour in the intermediate region.
If the eigenfunction decreases exponentially with depth in this region 
the mode has its largest amplitude in the outer region of the star and it
may be said to be {\it p-dominated}.
In the opposite case, with an eigenfunction increasing with depth, the
amplitude is largest in the core and the mode is {\it g-dominated}.
These two conditions essentially require the frequency to be determined
such that the mode resonates with
the outer acoustic, and the inner buoyancy-driven, cavity, respectively.
For the p-dominated modes this leads to Eq.~(\ref{eq:duvall}) for the
frequencies, which may be cast, at least approximately, in the form of
Eq.~(\ref{eq:tass}).
For the g-dominated modes we similarly find that the frequencies satisfy
Eqs~(\ref{eq:gfreq}) or (\ref{eq:gper}).
For evolved red giants like Model 3, with a very high buoyancy frequency in
the core, the period spacing $\Pi_0/L$ (cf.\ Eq~\ref{eq:gpersp})
becomes very small and hence the density
of the g-dominated modes is much higher than the density of p-dominated modes,
whose frequency spacing is given by $\Delta \nu$.
It should also be noted that the high density of g-dominated modes 
correspond to an extremely high number of nodes in the eigenfunctions in the
core, requiring substantial care in the numerical computation of the
oscillations.

\begin{figure}
\begin{center}
\includegraphics[width=12.0cm]{\fig/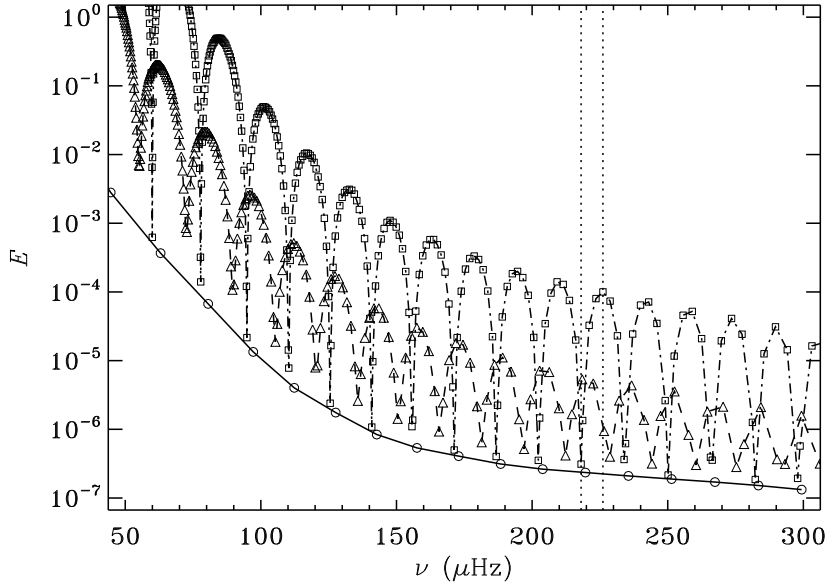}
\caption{Dimensionless inertia (cf.\ Eq~\ref{eq:inertia})
for Model~3 in the $1 \, M_\odot$ evolution sequence illustrated in
Fig.~\ref{fig:hr1}.
Radial modes are shown with circles connected by solid lines,
$l = 1$ modes with triangles connected by dashed lines and
$l = 2$ modes with squares connected by dot-dashed lines.
The vertical dotted lines mark the $l = 2$ modes 
for which the eigenfunctions are illustrated in Fig.~\ref{fig:eigenf}.
}
   \label{fig:inertia}
\end{center}
\end{figure}

\begin{figure}
\begin{center}
\includegraphics[width=8.0cm]{\fig/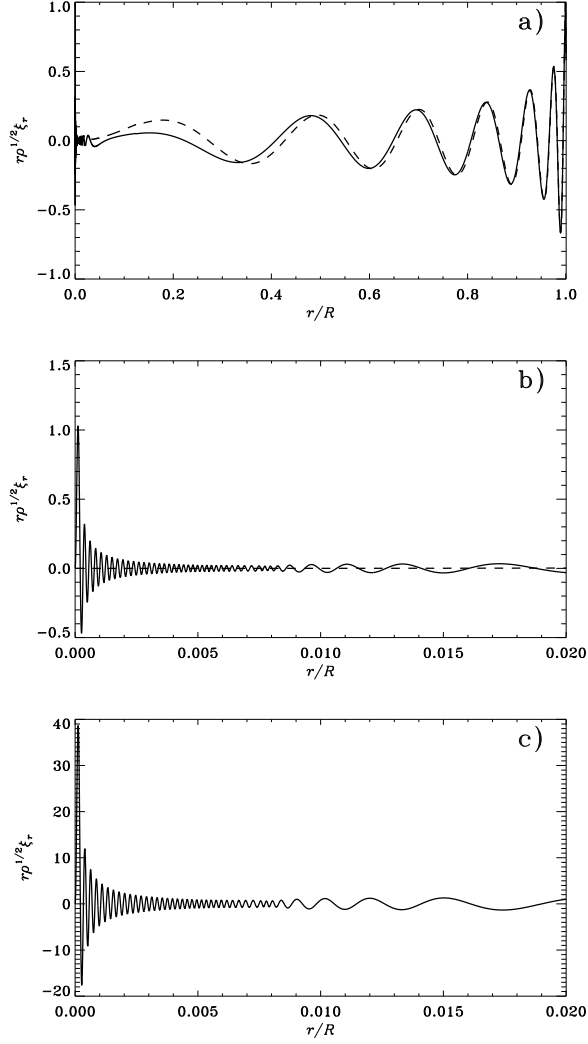}
\caption{Scaled eigenfunctions, normalized at the surface,
for Model~3 in the $1 \, M_\odot$ evolution
sequence shown in Fig.~\ref{fig:hr1}.
The quantity shown is $r \rho^{1/2} \xi_r$, where $\xi_r$ is the amplitude
of the vertical displacement.
In panels a) and b) the
dashed line shows a radial mode with cyclic frequency $\nu = 219.6 \muHz$ and
mode inertia (cf.\ Eq.~\ref{eq:inertia})
$E = 2.35 \times 10^{-7}$, and
the solid line shows a p-dominated $l = 2$ mode with
$\nu = 218.1 \muHz$ and $E = 3.12 \times 10^{-7}$.
In panel c) is illustrated a neighbouring g-dominated
$l = 2$ mode with $\nu = 226.1 \muHz$ and $E = 1.00 \times 10^{-4}$.
}
   \label{fig:eigenf}
\end{center}
\end{figure}

The properties of the modes can be illustrated by considering the normalized
mode inertia
\begin{equation}
E = {\int_V \rho |\bolddelr|^2 \dd V \over
M |\bolddelr|_{\rm s}^2} \; ,
\label{eq:inertia}
\end{equation}
where the integral is over the volume $V$ of the star and
$|\bolddelr|_{\rm s}^2$ is the surface value of the squared norm
of the displacement vector $\bolddelr$.
This is shown in Fig.~\ref{fig:inertia} for modes of degree $l = 0 - 2$
in Model 3.
The radial modes ($l = 0$) are purely acoustic, with an inertia that decreases
with increasing frequency.
For $l = 1$ and $2$, on the other hand, 
there is a dense set of g-dominated modes,
with acoustic resonances where the inertia approaches the value for the
radial modes at the corresponding frequency.
The pattern of these resonances is very similar to Eq.~(\ref{eq:tass}),
with the $l = 1$ resonances roughly half-way between the $l = 0$ 
frequencies and a small separation between an $l = 2$ resonance and the
neighbouring $l = 0$ mode.

Figure~\ref{fig:eigenf} shows a few selected eigenfunctions, in the
form of scaled radial component of the displacement, normalized to unity
at the surface.
The comparison between the neighbouring $l = 0$ and 2 modes in panel a)
shows that they are very similar in the bulk of the star where the $l = 2$
mode is predominantly of acoustic character.
In the core the eigenfunction for $l = 2$ varies rapidly but at generally small
amplitude for this p-dominated mode, as seen in panel b).
The behaviour in the core for the mode shown in panel c) 
is very similar, but at a
much higher amplitude relative to the surface normalization, and
corresponding to the local maximum in the inertia.
We discuss the consequences for the observable visibility of the modes in
Section~\ref{sec:rgamplitude} below.


\section{Damping and excitation of solar-like oscillations}

\label{sec:excitation}
When nonadiabatic effects are taken into account, the oscillation frequencies
are no longer real. 
Assuming again a time dependence as $\exp( - \eye \omega t)$ and writing
$\omega = \omega_{\rm r} + \eye \omega_{\rm i}$ in terms of real and imaginary
parts,
the real part $\omega_{\rm r}$ determines the frequency of the mode whereas
the imaginary part $\omega_{\rm i}$ determines the growth or decay 
of the amplitude, $\omega_{\rm i} < 0$ corresponding to a damped mode.
The determination of the damping or excitation of the modes requires 
a solution of the full nonadiabatic oscillation equations, introducing also
the energy equation and equations for the energy generation and transport.
The latter is given by the perturbations to the flux
$\boldF = \boldF_{\rm rad} + \boldF_{\rm con}$ which, as indicated,
has both a radiative ($\boldF_{\rm rad}$) and convective ($\boldF_{\rm con}$) 
contribution.
The radiative part can be dealt with in a relatively straightforward manner, 
while the treatment of the convective part remains highly uncertain.
Furthermore, a significant contribution to the damping may come from
the dynamical effects of convection through the perturbation to the Reynolds
stresses, often approximated by a turbulent pressure $p_{\rm t}$,
which again is uncertain \citep[e.g.,][]{Houdek2010a}.

A convenient way to discuss the damping or excitation of the modes
is to express the growth rate, approximately in terms of the 
{\it work integral} $W = W_{\rm g} + W_{\rm t}$, as
\begin{equation}
\omega_{\rm i} 
\simeq {W_{\rm g} \over J} + {W_{\rm t} \over J} \; ,
\label{eq:wi}
\end{equation}
with
\begin{eqnarray}
W_{\rm g} &=& {\rm Re} \left[ \int_V {\delta \rho^* \over \rho} (\Gamma_3 - 1)
\delta (\rho \epsilon - \div \boldF) \dd V \right] \; , \nonumber \\
W_{\rm t} &=& - \omega_{\rm r}{\rm Im} \left[ \int_V {\delta \rho^* \over \rho} 
\delta p_{\rm t} \dd V \right] \; , \nonumber \\
J &=& 2 \omega_{\rm r}^2 \int_V \rho |\bolddelr|^2 \dd V 
= 2 \omega_{\rm r}^2 M |\bolddelr_{\rm s}|^2 E \; .
\label{eq:wiparts}
\end{eqnarray}
Here $\delta$ denotes the Lagrangian perturbation, i.e., the perturbation
following the motion,
$\Gamma_3 - 1 = (\partial \ln T/ \partial \ln \rho)_{\rm ad}$, 
and $\epsilon$ is the rate of energy generation per unit mass;
also the star indicates the complex conjugate.

Equations (\ref{eq:wi}) and (\ref{eq:wiparts}) have a simple physical meaning:
In the expression for $W_{\rm g}$ $\delta (\rho \epsilon - \div \boldF)$
gives the perturbation to the heating rate;
thus $W_{\rm g}$ gets a positive contribution,
and hence a contribution to the excitation of the mode,
where heating and compression, described by $\delta \rho/\rho$, are in phase.
Heating at compression is the basis for the operation of any heat engine,
and hence modes that are unstable because of this can be said to be driven by
the {\it heat-engine} mechanism.%
\footnote{Since the perturbation to the heating is often determined by the
perturbation to the opacity $\kappa$ this is also known as the 
$\kappa$ mechanism.}
Depending on the phase relation between compression and $\delta p_{\rm t}$
the term in $W_{\rm t}$ may contribute to the damping or driving of the mode.

As already indicated, the treatment of the convection-pulsation interaction is
a major uncertainty in nonadiabatic calculations for stars with significant
outer convection zones, where the convective timescales and the pulsation 
periods are typically similar.
Two generalizations of mixing-length theory have seen fairly widespread use.
In one, based on an original model by \citet{Unno1967}, steady convective eddies
are considered, with a balance between buoyancy and turbulent viscous drag.
In the second, developed by \citet{Gough1977}, convective eddies are continually
created and destroyed.
The effects of the perturbations associated with the pulsations on these
physical models can then be described, leading to expressions for the
perturbations to the convective flux and Reynolds stresses.
Unno's description has been further developed 
\citep[see][and references therein]{Grigah2005},
while Gough's description was developed and applied by,
e.g., \citet{Balmfo1992} and \citet{Houdek1999}.
With judicious (but not unreasonable) choices of parameters both formulations
can predict the transition to stability at the red edge of the Cepheid 
instability strip, as well as yield results for solar models that 
are consistent with the observationally inferred damping rates.%
\footnote{A different formulation by \citet{Xiong1997}, based 
on equations for the second- and third-order correlations,
similarly predicts the red edge but apparently finds that some of the observed
solar modes are unstable \citep{Xiong2010}.}
Interestingly, the dominant mechanism resulting in damping of the modes
differs between the two formulations: in the calculations based on Gough's 
formulation damping is dominated by the perturbation to the turbulent pressure,
while the results based on Unno's formulation show damping through the
perturbation to the convective flux.
A more detailed comparison of these two sets of results is clearly called for.

\begin{figure}
\begin{center}
\includegraphics[width=8.0cm]{\fig/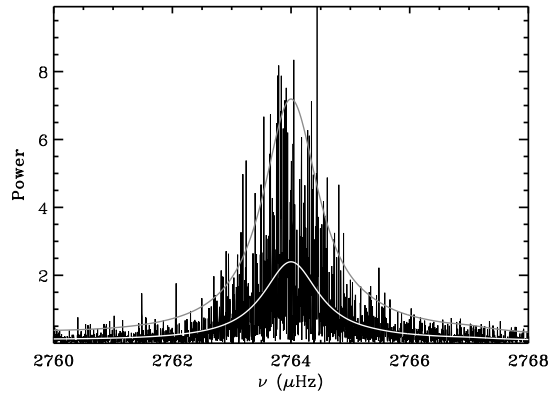}
\caption{Power spectrum of a solar radial mode, from 8 years of BiSON
observations.
The white curve shows a Lorentzian fit (cf.\ Eq.~\ref{eq:stochpow})
to the observed spectrum, while the smooth grey curve shows the same
fit, but multiplied by a factor three.
Data courtesy of W. J. Chaplin; see \citet{Chapli2002}.
}
   \label{fig:bisonfit}
\end{center}
\end{figure}

The cause of the solar oscillations is now almost universally thought to be
stochastic excitation by near-surface convection of the intrinsically stable
modes.
Stochastic excitation of waves in the solar atmosphere was first considered
by \citet{Stein1968}, followed by the application to the excitation of 
normal modes by \citet{Goldre1977}.
The basic properties of damped oscillations excited by turbulence was
considered by \citet{Batche1956} \citep[see also][]{Christ1989}.
The result is that the power spectrum of a single mode, of natural frequency
$\omega_0$ and damping rate $\omega_{\rm i}$, is 
\begin{equation} 
P(\omega) \simeq {1 \over 4 \omega_0^2} {P_{\rm f}(\omega) \over 
(\omega - \omega_0)^2 + \omega_{\rm i}^2} \; ,
\label{eq:stochpow}
\end{equation}
where $P_{\rm f}(\omega)$ is the power spectrum of turbulent forcing.
Assuming that $P_{\rm f}$ varies slowly with frequency the result is a
stochastic function modulated by a Lorentzian envelope, with a full 
width at half maximum of $2 |\omega_{\rm i}|$.
An example, from BiSON observations of a solar radial mode, is shown in
Fig.~\ref{fig:bisonfit}.
{}From the fitted Lorentzian it is obviously possible to determine the
damping rate $|\omega_{\rm i}|$.
Further analyses of the statistical properties of the stochastically
excited oscillators were carried out by \citet{Kumar1988} and \citet{Chang1998}.
It was shown by \citet{Chapli1997} that the observed distribution of solar
oscillation amplitudes was consistent with the predictions of these analyses.

An early rough prediction of the excitation of solar-like modes in other stars
was carried out by \cite{Christ1983}.
It was shown by \citet{Kjelds1995} that the resulting velocity amplitudes
approximately scaled as $L/M$, where $L$ is the surface luminosity of the star.
Later, more careful investigations
\citep[e.g.,][]{Houdek1999, Samadi2007} confirm that the amplitudes 
increase with $L/M$ although generally to a power slightly less than one.

The excitation of a specific mode clearly depends on the detailed property
of the mode,
the amplitude being determined by the balance between
the stochastic energy input from convection and the damping.
The energy input increases as a high power of the convective
velocity \citep{Stein1968}
and hence is dominated by the near-surface layers where convection is
most vigorous.
Here the properties of the oscillations depend predominantly on the frequency,
with little dependence on the degree (at least for the low-degree modes
that are relevant in distant stars).
The outcome is that the mean square amplitude can be written as
\citep[e.g.,][]{Chapli2005}
\begin{equation}
\langle A^2 \rangle \simeq {1 \over E |\omega_{\rm i}|} 
{{\cal P}_{\rm f}(\omega) \over E} \; ,
\label{eq:amplitude}
\end{equation}
where $A$ might be the surface velocity or the relative intensity 
fluctuation;
here ${\cal P}_{\rm f}(\omega)$, which depends on frequency but 
in general not on degree, describes the energy input from convection
\citep[see, e.g.,][for a review]{Houdek2010b}.
Also, the damping rate is given by Eqs~(\ref{eq:wi}) and (\ref{eq:wiparts}).
In the common case where the integrals in $W_{\rm g}$ and $W_{\rm t}$
are dominated by the near-surface layers it follows that 
$|\omega_{\rm i}| E$ is independent of $E$ at constant frequency;
consequently, in this case $\langle A^2 \rangle \propto E^{-1}$
at constant frequency.
This is important in the discussion of the excitation of 
solar-like oscillations in red giants where we see large variations in 
$E$ over a narrow frequency range (cf.\ Fig.~\ref{fig:inertia}):
the g-dominated modes, with large inertias,
are clearly expected to have a much smaller root-mean-square amplitude 
than the p-dominated modes with low inertia.
However, the question of the mode {\it visibility}, and hence
the likelihood to detect them, is more complicated;
we return to this in Section~\ref{sec:rgamplitude}.

The balance between energy input and damping
leads to a characteristic bell-shaped amplitude distribution of
the mean amplitude, although with fluctuations around the mean reflecting
the stochastic nature of the excitation.
The shape of the amplitude distribution was discussed by \citet{Goldre1994}.
They found that the increase in amplitude with frequency at low frequency
was predominantly caused by the change in the shape of the eigenfunction
as the upper reflection point, where $\omega = \omega_{\rm c}$ moves 
closer to the surface with increasing $\omega$.
At high frequency this upper reflection point is very close to the
photosphere, and the amplitude decreases with increasing frequency
due to an increasing mismatch between the oscillation periods and
the timescales of the convective eddies that dominate the excitation.
As the frequency approaches $\omega_{\rm c, atm}$ beginning energy loss 
through the stellar atmosphere may also contribute to the damping and
hence further decrease the amplitudes.
Indeed, the variation with frequency of the damping rate clearly also
plays a role for the the variation in amplitude.

It has been found observationally that the cyclic frequency
$\nu_{\rm max}$ of maximum amplitude appears to scale with 
the acoustic cut-off frequency $\nu_{\rm c, atm}$
\citep[e.g.,][]{Brown1991, Kjelds1995, Beddin2003, Stello2008}
(see also Bedding, this volume),
and hence follows the scaling in Eq.~(\ref{eq:cutoff}).
This scaling has been a very useful tool in characterizing
the observed properties of
stars showing solar-like oscillations \citep[e.g.,][]{Kallin2010},
but the relation between $\nu_{\rm max}$ and $\nu_{\rm c, atm}$
still defies a definite theoretical understanding.
\citet{Belkac2011} noted that the location of $\nu_{\rm max}$ is closely
related to the plateau in damping rates at intermediate frequency,
apparently caused by a resonance between the pulsation period and the local
thermal timescale,
a feature already noticed by \citet{Gough1980} and \citet{Christ1989}.
However, this does not fully explain the observed tight relation.


\section{Do red giants have nonradial oscillations?}

%
\label{sec:rgnonradial}
As pointed out by \citet{Kjelds1995} the amplitude predictions by
\citet{Christ1983} scaled as $L/M$.
Thus one could expect to see solar-like oscillations in red giants with
quite substantial amplitudes.
Early evidence was found for such oscillations in Arcturus
\citep{Smith1987, Innis1988}
while \citet{Edmond1996} identified oscillations in K giants in the
globular cluster 47 Tuc.
\citet{Christ2001} noted that the statistics of very long-period semi-regular
variables observed by the Americal Association of Variable Star Observers
appeared to obey amplitude statistics consistent with stochastic excitation,
strongly suggesting that the variability of these stars was also solar-like,
although at month-long periods and with amplitudes allowing simple visual
observations.
Photometric observations of the oscillations in Arcturus
were carried out by \citet{Retter2003} with the WIRE satellite.
Analyses of large samples of red giants observed by OGLE%
\footnote{Optical Gravitational Lensing Experiment}
\citep[e.g.][]{Kiss2003, Kiss2004, Soszyn2007} 
showed oscillations in several modes, 
and \citet{Dziemb2010} confirmed that the most likely cause was solar-like
oscillations.
In a further analysis of ground-based survey observations of red giants
in the Galaxy and the Large Magellanic Could \citet{Tabur2010} 
demonstrated a continuous transition between pulsations in G and K giants
and those seen in very luminous M giants.

\begin{figure}
\begin{center}
\includegraphics[width=10.0cm]{\fig/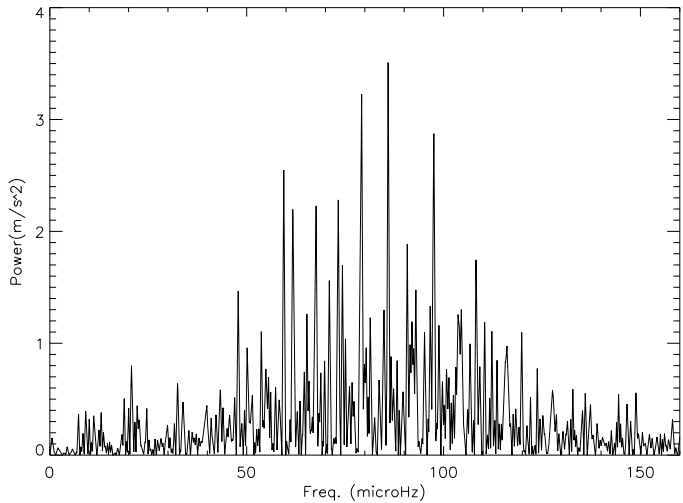}
\caption{Power spectrum of the G7 giant $\xi$ Hydrae, from one month of
Doppler velocity measurements.
Adapted from \citet{Frands2002}.
}
   \label{fig:xhyaspec}
\end{center}
\end{figure}

The first detailed observations of the oscillations of a red giant was carried
out by \citet{Frands2002} for the G7 giant $\xi$ Hya.
The resulting power spectrum is shown in Fig.~\ref{fig:xhyaspec}.
This clearly shows peaks at uniformly spaced frequency, in accordance with
Eq.~(\ref{eq:tass}), and a power envelope very similar to what is observed
in the Sun, strongly supporting the solar-like nature of the oscillations.
{}From analysis of the spectrum the separation between the peaks 
was determined to be $\simeq 7 \muHz$.

\begin{figure}
\begin{center}
\includegraphics[width=8.0cm]{\fig/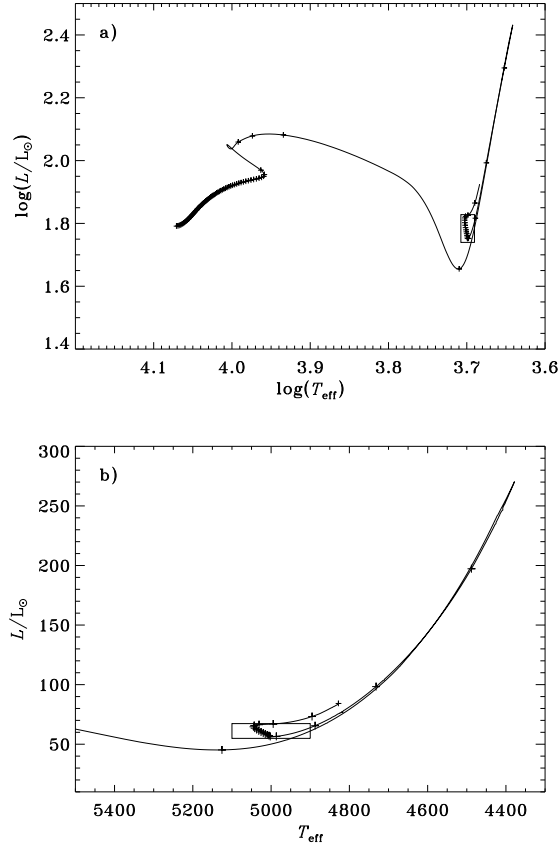}
\caption{Evolution track for a model of $\xi$ Hydrae,
with mass $M = 2.8 M_\odot$ and heavy-element abundance $Z = 0.019$.
The plusses are placed at 5\,Myr intervals along the track.
The box marks the $1{-}\sigma$ error box for the observed effective temperature
and luminosity.
}
   \label{fig:xhyahr}
\end{center}
\end{figure}

Modelling of $\xi$ Hya to match the well determined luminosity and effective
temperature is shown in Fig.~\ref{fig:xhyahr}.
Stars ascend the red-giant branch while burning hydrogen in a shell around
an inert and almost isothermal helium core.
As the mass of the core increases, so does the luminosity and the temperature
of the core. 
At the tip of the red-giant branch the temperature reaches a sufficiently
high value, around $10^8 \, {\rm K}$, where helium fusion becomes significant.
The resulting change in structure, with an expansion of the core,
leads to a reduction in the surface radius and luminosity, with the
star entering a phase of stable helium burning.
Even during this phase, however, a substantial part of the total
energy production takes place in the hydrogen-burning shell.
As the phase is relatively extended, with a duration of around 20 per cent of
that of the central hydrogen-burning main sequence, the chances of observing 
a star in this phase are fairly high;
in stellar clusters such stars populate the so-called `red clump'.

As noted by \citet{Teixei2003} $\xi$ Hya could be identified as being in
the ascending phase on the red giant branch.
However, since this phase is extremely brief a more likely identification would
be with the core helium burning phase.
In either case, a match to the observed frequencies was possible only if 
the observed frequency spacing was identified with $\Delta \nu$, such that only
modes of one degree, presumably radial modes, were observed.
In contrast, from a strict application of Eq.~(\ref{eq:tass})
one would have expected the
spacing to be $\Delta \nu/2$, and hence $\Delta \nu \simeq 14 \muHz$.
However, given the scaling in Eq.~(\ref{eq:nuscale}) this would be entirely
inconsistent with the location of the star in the HR diagram.

To account for this identification
\citet{Christ2004} considered the possibility that all nonradial modes
in red giants would likely be heavily damped and hence have small amplitudes.
He noted that $W_{\rm g}$ (cf.\ Eq.~\ref{eq:wiparts})
depends on $\delta (\div \boldF)$, i.e., on the derivative of $\delta \boldF$.
In turn, roughly $\delta \boldF \propto \nabla (\delta T)$ and
$\delta T/T \propto \delta \rho/\rho \propto \div \bolddelr$.
It follows that $W_{\rm g}$ depends on the third derivative of 
$\bolddelr$. 
Thus even if the amplitude of $\bolddelr$ in the core is quite small,
as is the case for p-dominated modes (cf.\ Fig.~\ref{fig:eigenf}),
$\delta (\div \boldF)$ could be quite large for a mode varying very
rapidly as a function of $r$, leading to a substantial contribution to the
damping from the core for all nonradial modes.
Although the analysis was entirely qualitative,
\cite{Christ2004} found some support for this conjecture in the analyses
by \citet{Dziemb1977} and \citet{Dziemb2001}.

\begin{figure}
\begin{center}
\includegraphics[width=6.0cm]{\fig/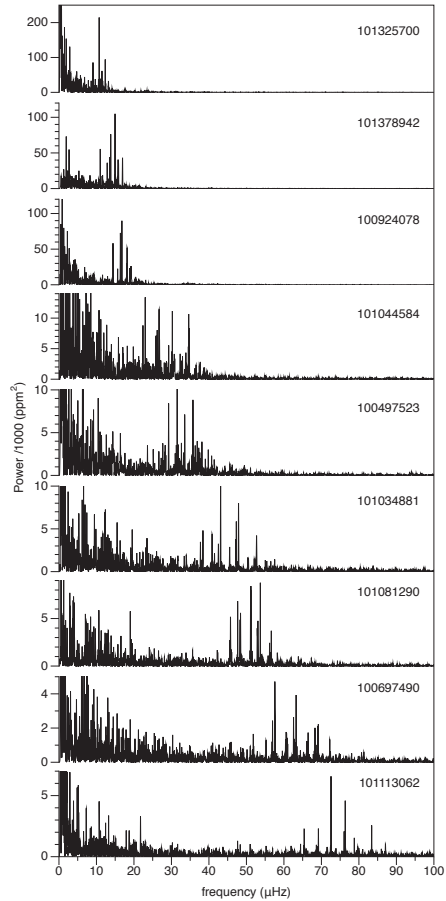}
\caption{Power spectra of solar-like oscillations in red giants,
from 5 months of observations with the CoRoT satellite.
The panels are labelled by the CoRoT identification number.
Radius and luminosity increase towards the top.
Figure courtesy of J. De Ridder; from \citet{DeRidd2009}.
}
   \label{fig:corotrg}
\end{center}
\end{figure}

\begin{figure}
\begin{center}
\includegraphics[width=8.0cm]{\fig/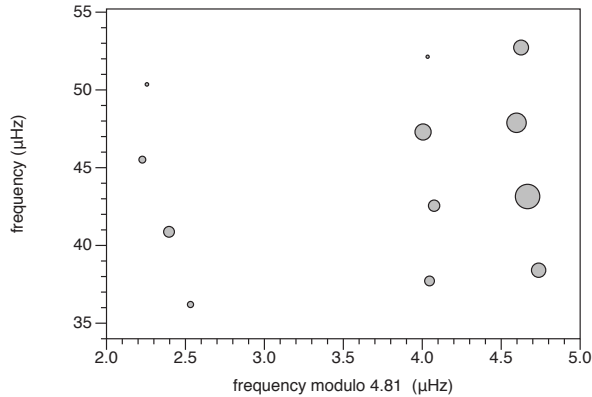}
\caption{\'Echelle diagram of the star 101034881 in Fig.~\ref{fig:corotrg}.
This corresponds to stacking slices of the power spectrum of
a length given the large separation $\Delta \nu = 4.81 \muHz$:
the abscissa is the frequency modulo $\Delta \nu$ whereas the ordinate is the
frequency.
Figure courtesy of J. De Ridder; from \citet{DeRidd2009}.
}
   \label{fig:corotechl}
\end{center}
\end{figure}

This pessimistic expectation for red-giant seismology has been completely
overturned by observations and an improved theoretical understanding.
\citet{Hekker2006} found evidence for nonradial pulsation in the line-profile
variation of a few red giants showing solar-like oscillations.
Even more dramatically, \citet{DeRidd2009} clearly demonstrated the presence
of nonradial solar-like oscillations in a large number of red giants 
from early CoRoT observations.
Figure~\ref{fig:corotrg} shows resulting power spectra.
The power enhancement from the solar-like oscillations is evident;
as stars on the red-giant branch have roughly the same effective
temperature,
the frequency at maximum power decreases with increasing radius
and luminosity in accordance with Eq.~(\ref{eq:cutoff}).
The observed frequencies satisfy a relation similar to Eq.~(\ref{eq:tass}),
with definite evidence for nonradial modes.
This is most clearly illustrated in an {\it \'echelle diagram}
(see Bedding, this volume),
where the frequencies are shown against the frequency
reduced modulo $\Delta \nu$, effectively dividing the spectrum into
segments of length $\Delta \nu$ and stacking the segments.
This has been done for one of the stars in Fig.~\ref{fig:corotechl}.
In accordance with Eq.~(\ref{eq:tass}) the three columns of points
correspond to modes of degree $l = 0$ and $2$ to the right, separated
by the small separation, and modes with $l = 1$ to the left.
Analyses of red giants observed by CoRoT and {\it Kepler}
\citep[e.g.,][]{Huber2010, Mosser2011a} 
have shown that this frequency pattern is universal amongst red giants.

The very long time series obtained by CoRoT and in particular {\it Kepler}
have allowed fits to the power spectra in terms of the Lorentzian
profiles (cf.\ Eq.~\ref{eq:stochpow}) and hence a determination
of the mode damping rates or, equivalently, the lifetimes
$\tau = 1/|\omega_{\rm i}|$.
For red giants this yields lifetimes near the maximum amplitude
of around 15 days \citep{Huber2010, Baudin2011}.
Interestingly, this is very similar to the value obtained
for $\xi$ Hya by \citet{Houdek2002},
using a convection treatment based on \citet{Gough1977}.

\section{Amplitudes of solar-like oscillations in red giants}

\label{sec:rgamplitude}
%
To interpret the observations a more detailed analysis of the mode excitation
and visibility is required.
The analysis leading to Eq.~(\ref{eq:amplitude}) determines 
the mean square amplitude of the mode, corresponding essentially
to the area under the corresponding peak in the power spectrum.
However, it was noted by \citet{Chapli2005} that ability to detect
a mode in the power spectrum depends instead on the {\it peak height} $H$,
related to $A^2$%
\footnote{for simplicity, we drop $\langle ... \rangle$ in the following}
by
\begin{equation}
A^2 = |\omega_{\rm i}| H \; ,
\label{eq:heightinf}
\end{equation}
since $|\omega_{\rm i}|$ determines the width of the peak.
If the damping is dominated by the near-surface layers, it was argued
above that both $A^2$ and $|\omega_{\rm i}|$ are proportional 
to $E^{-1}$ at a given frequency.
It then follows from Eq.~(\ref{eq:heightinf}) that $H$ is independent
of $E$, i.e., that all modes in a given frequency interval are excited
to roughly the same amplitude.
For the dense spectrum of modes in a red giant, such as illustrated in
Fig.~\ref{fig:inertia}, the result would be a very dense 
observed power spectrum, with no indication of the acoustic resonances
approximately satisfying Eq.~(\ref{eq:tass}).
This is in obvious contradiction to, e.g., the observations
by \citet{DeRidd2009}.

This argument, however, neglects the fact that the oscillations are observed
for a finite time $\CT$ which, for the g-dominated modes,
is typically shorter than the lifetime $\tau$ of the mode.
The resulting peaks in the power-density spectrum has a width that, for 
$\CT \ll \tau$, is proportional to $\CT^{-1}$ 
and hence a correspondingly smaller peak height,
very likely making the modes invisible.
As a convenient interpolation between the two extremes, \citet{Fletch2006}
proposed the relation
\begin{equation}
H \propto {A^2 \over |\omega_{\rm i}| + 2/\CT} \; .
\label{eq:peakheight}
\end{equation}

\begin{figure}
\begin{center}
\includegraphics[width=10.0cm]{\fig/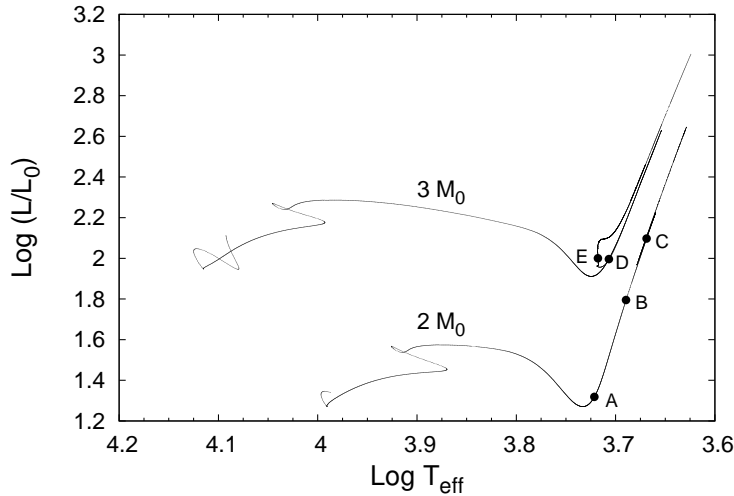}
\caption{Evolution tracks for models of mass 2 and $3 \, M_\odot$,
for which \citet{Dupret2009} considered the stochastic excitation of
solar-like oscillations. Figure courtesy of M.-A. Dupret.
}
   \label{fig:dup1}
\end{center}
\end{figure}

A detailed and very illuminating analysis of the excitation of modes in red
giants was carried out by \citet{Dupret2009}. 
They considered several red-giant models, with locations in the HR diagram
as illustrated in Fig.~\ref{fig:dup1}, carrying out nonadiabatic calculations
of nonradial modes using the convection formalism of \citet{Grigah2005}.
{}From a model of the stochastic energy input 
\citep[e.g.,][]{Belkac2006} they then computed the 
expected amplitudes of the modes and the resulting power-density diagrams.
They assumed an observing period $\CT$ of 150 days,
corresponding to CoRoT long runs.
Rather than using Eq.~(\ref{eq:peakheight}) they effectively adopted
$H \propto \tau A^2$ for $\tau > \CT/2$ and $H \propto \CT A^2/2$ otherwise,
with a continuous transition between the two cases.
Here we briefly consider their results for Models A and B.

\begin{figure}
\begin{center}
\includegraphics[width=8.0cm]{\fig/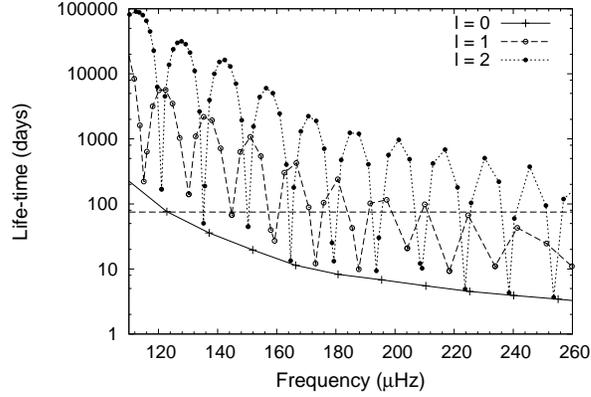}
\caption{Computed lifetimes of Model~A in Fig.~\ref{fig:dup1},
at the base of the $2 M_\odot$ red-giant branch.
Modes of degree $l = 0$, 1 and 2 are shown, respectively,
by plusses connected by
solid lines, open circles connected by dashed lines and 
closed circles connected by dotted lines.
The horizontal dashed line marks a damping time of 75\,d, corresponding to the
border between unresolved and resolved modes for a CoRoT long run of 150 days.
Figure courtesy of M.-A. Dupret; from \citet{Dupret2009}.
}
   \label{fig:dup5a}
\end{center}
\end{figure}

\begin{figure}
\begin{center}
\includegraphics[width=8.0cm]{\fig/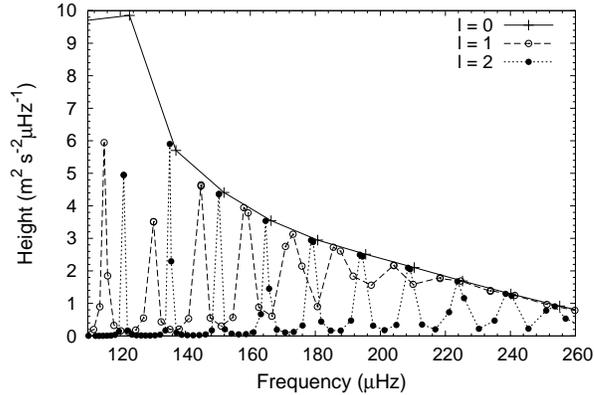}
\caption{Estimated peak heights in the power spectrum of
solar-like oscillations in Model~A in Fig.~\ref{fig:dup1},
at the base of the red-giant branch;
for symbol and line styles, see caption to Fig.~\ref{fig:dup5a}.
Figure courtesy of M.-A. Dupret; from \citet{Dupret2009}.
}
   \label{fig:dup5d}
\end{center}
\end{figure}

Figure \ref{fig:dup5a} shows the mode lifetimes for modes of degree $l = 0 - 2$
in Model~A.
In this case the damping is dominated by the near-surface layers and hence, as
is obvious in the figure, the variation of $\tau$ reflects the behaviour
of $E$ which is qualitatively similar to the case illustrated in
Fig.~\ref{fig:inertia}.
For $l = 2$ the g-dominated modes have lifetimes substantially higher than the
assumed $\CT/2$, indicated in the figure by a horizontal dashed line.
On the other hand, the $l = 1$ modes at higher frequency all have lifetimes
at or below the observing time, and hence would be expected to reach similar
peak heights in a given frequency interval.
This is confirmed by Fig.~\ref{fig:dup5d} which shows the predicted 
peak heights.
The overall envelope seems to be somewhat skewed towards lower frequency,
relative to what is expected from the observed scaling with the acoustic
cut-off frequency;
given the uncertainties in modelling both the damping and the energy input,
this is hardly surprising.
However, it is striking that while the $l = 2$ modes show strong
acoustic resonances, with small peak heights for the g-dominated modes,
the $l = 1$ modes at high frequency all reach peak heights comparable
with those for the radial modes.
It is obvious that such a spectrum can be quite complex
\citep[see also][]{DiMaur2011}.

\begin{figure}
\begin{center}
\includegraphics[width=8.0cm]{\fig/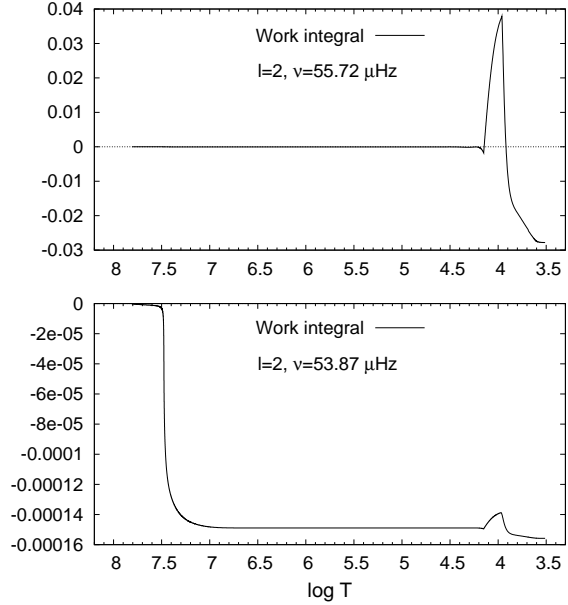}
\caption{Work integrals for the intermediate red-giant branch 
$2 \,M_\odot$ Model~B in Fig.~\ref{fig:dup1};
this essentially corresponds to the partial integrals 
of $W_{\rm g} + W_{\rm t}$, defined in Eq.~(\ref{eq:wiparts}),
and here plotted against temperature $T$ in the model.
The top panel is for a p-dominated mode while the bottom panel is for
a g-dominated mode largely trapped in the core.
Figure courtesy of M.-A. Dupret; from \citet{Dupret2009}.
}
   \label{fig:dup8}
\end{center}
\end{figure}

\begin{figure}
\begin{center}
\includegraphics[width=8.0cm]{\fig/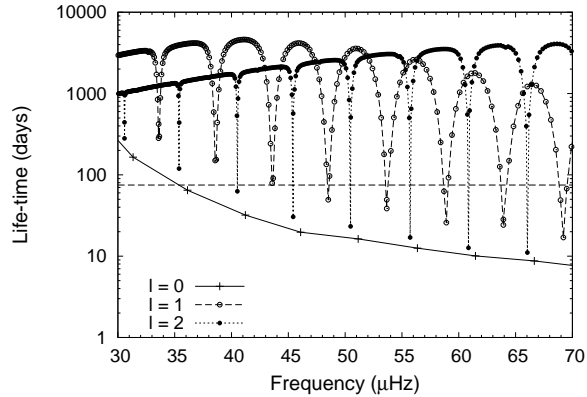}
\caption{Computed lifetimes for the intermediate red-giant branch 
$2 \,M_\odot$ Model~B in Fig.~\ref{fig:dup1};
see caption to Fig.~\ref{fig:dup5a}.
Figure courtesy of M.-A. Dupret; from \citet{Dupret2009}.
}
   \label{fig:dup7a}
\end{center}
\end{figure}

\begin{figure}
\begin{center}
\includegraphics[width=8.0cm]{\fig/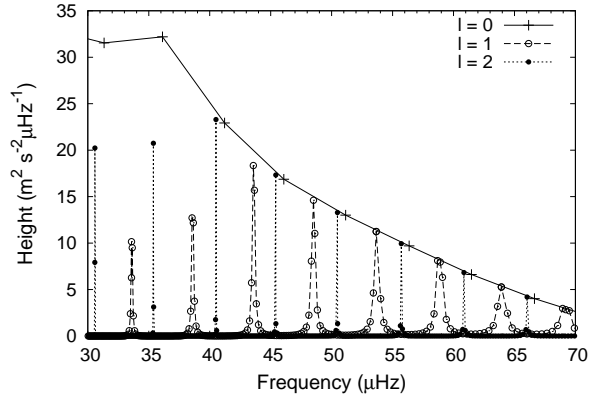}
\caption{Estimated peak heights in the power spectrum of
solar-like oscillations in the intermediate red-giant branch 
$2 \,M_\odot$ Model~B in Fig.~\ref{fig:dup1};
for symbol and line styles, see caption to Fig.~\ref{fig:dup5a}.
Figure courtesy of M.-A. Dupret; from \citet{Dupret2009}.
}
   \label{fig:dup7d}
\end{center}
\end{figure}

The situation is rather different for the more evolved Model~B.
As illustrated in Fig.~\ref{fig:dup8} there is now strong damping in the
core for the g-dominated modes with $l = 2$;
for $l = 1$ the same is the case at relatively low frequency, while
the surface layers dominate at high frequency.
The damping in the core is a result of the very rapidly oscillating 
eigenfunctions in the core, as discussed above.
Owing to the increased central condensation the buoyancy frequency 
is substantially higher than in Model~A, leading to a higher vertical wave
number and hence stronger damping in the core.
For p-dominated modes, on the other hand, the damping is still dominated 
by the near-surface layers.
Together with a strong trapping of the g-dominated modes in the core,
this leads to a much stronger contrast in lifetime between the p- and
g-dominated modes, with the latter having lifetimes very substantially
higher than the observing time (cf.\ Fig.~\ref{fig:dup7a}).
The effect of this on the predicted power density spectrum is evident in
Fig.~\ref{fig:dup7d}:
there are now obvious acoustic resonances for both $l = 1$ and 2.
The $l = 2$ resonances are confined to very narrow frequency
intervals, caused also by the more extended evanescent region
between the peak in the buoyancy frequency and the region of acoustic
propagation, and typically involve only
1 -- 2 modes at non-negligible peak heights,
For $l = 1$, on the other hand,
there is typically a cluster of peaks with heights that may make the
modes observable.
As discussed by Bedding (this volume) this was found to be the case
by \citet{Beck2011} and \citet{Beddin2011}.
We discuss the significance of these results in the following section.

\section{Asteroseismic diagnostics in red giants}

%
\label{sec:rgastero}
Detection of solar-like oscillations in red giants, such as illustrated
in Fig.~\ref{fig:corotrg}, almost immediately provides a measure
of the large frequency separation $\Delta \nu$ and the frequency 
$\nu_{\rm max}$ of maximum power.
These basic quantities already have a substantial diagnostic potential:
$\Delta \nu$ approximately satisfies the scaling (Eq.~\ref{eq:nuscale}) with
the mean density and it can be assumed that $\nu_{\rm max}$ scales
as the atmospherical acoustic cut-off frequency (Eq.~\ref{eq:cutoff}).
If $T_{\rm eff}$ is known from photometric or spectroscopic observations
one can then determine the radius and mass of the star as
\begin{equation}
{R \over R_\odot} = {\nu_{\rm max} \over \nu_{\rm max,\odot}}
\left({\Delta \nu \over \Delta \nu_\odot} \right)^{-2}
\left({T_{\rm eff} \over T_{\rm eff, \odot}} \right)^{1/2} 
\end{equation}
and
\begin{equation}
{M \over M_\odot} = \left({R \over R_\odot} \right)^3 
\left({\Delta \nu \over \Delta \nu_\odot} \right)^2 \; ,
\end{equation}
where $\odot$ denotes solar values \citep[e.g.,][]{Kallin2010}.
Such analyses provide unique possibilities for population studies
of field red giants \citep[e.g.,][]{Miglio2009, Mosser2010, Hekker2011}.
Stellar modelling provides further constraints on the relation between
$T_{\rm eff}$, $M$ and $R$ and hence a more precise determination
of stellar properties \citep{Gai2011},
as used in an analysis of red giants in two of the open clusters
in the {\it Kepler} field by \citet{Basu2011}.

\begin{figure}[thp]
\begin{center}
\includegraphics[width=8.0cm]{\fig/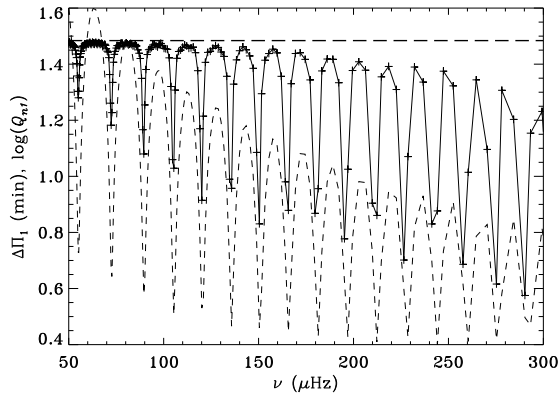}
\caption{The solid curve shows the period spacing between adjacent $l = 1$
modes in Model~3 in the $1 M_\odot$ sequence in Fig.~\ref{fig:hr1}.
The horizontal dashed line shows the asymptotic value for pure g modes,
from Eq.~(\ref{eq:perspac1}).
For comparison, the short-dashed curve shows the logarithm of the
normalized inertia $Q_{nl}$ (cf.\ Eq.~\ref{eq:qnl}) on an arbitrary scale.
}
   \label{fig:perspac}
\end{center}
\end{figure}

As discussed in connection with Eq.~(\ref{eq:smallsep}),
for main-sequence stars the small
frequency separations $\delta \nu_{l\,l+2}(n) = \nu_{nl} - \nu_{n-1\, l+2}$
provide a measure of the age of the star.
This no longer holds for red giants.
Here in most cases $\delta \nu_{l\,l+2}$ scales essentially as the large
frequency separation and hence mainly reflects the variation of the mean
density of the star \citep[e.g.,][]{Beddin2010, Huber2010, Mosser2011a}.
This is a natural consequence of the fact that the propagation region of the
nonradial acoustic modes for such stars lies in the convective envelope,
which probably changes in a largely homologous fashion as the properties of
the star change.
A detailed analysis of the diagnostic potential of
acoustic modes in red giants was carried out by \citet{Montal2010}.
They also considered other types of small frequency separations, such as
$\delta \nu_{01}(n) = (\nu_{n0} + \nu_{n+1\,0})/2 - \nu_{n1}$.
Interestingly, they found that for models with somewhat shallower 
convective envelopes, particularly amongst core helium-burning stars,
where the lower turning point of the $l = 1$ modes was in the radiative
region, the average $\langle \delta \nu_{01} \rangle/\Delta \nu$ showed a
marked dependence on the distance from the base of the convective envelope
to the turning point.

The asymptotic analysis leading to Eq.~(\ref{eq:tass}) for the acoustic-mode
frequencies assumes that the equilibrium structure varies slowly compared
with the eigenfunctions.
There are regions in the star where this does not hold, with variations
in the sound speed on a scale that is comparable with or smaller than the
local wavelength.
This causes periodic variations in the frequencies with properties that
depend on the depth and nature of the sound-speed feature.
In red giants the most important example of such an {\it acoustic glitch}
is related to the second ionization of helium which causes a dip in
$\Gamma_1$ and hence (cf.\ Eq.~\ref{eq:csqfull}) a localized variation
in the sound speed.
This variation provides a diagnostics of the helium abundance in the
star \citep[e.g.,][]{Gough1990, Voront1991, Basu2004, Montei2005, Houdek2007}.
A clear signature of this effect was found by \citet{Miglio2010} in a
red giant observed by CoRoT.
They noted, following an earlier discussion by \citet{Mazumd2005} of the
diagnostic potential of such glitches, that the inferred depth of the
sound-speed feature together with large frequency separation provide
a purely asteroseismic determination of the mass and radius of the star.
This feature can probably be detected in a substantial fraction of the
red giants observed with CoRoT and {\it Kepler},
promising a determination of the so far poorly known helium abundance in
these stars and hence important constraints on the chemical evolution of
the Galaxy.

From the analysis by \citet{Dupret2009}, discussed in 
Section~\ref{sec:rgamplitude}, we noted that for $l = 1$
several modes might be excited to observable amplitudes in the
vicinity of an acoustic resonance.
This is consistent with the rather larger scatter for these dipolar
modes observed by \citet{Beddin2010} in an collapsed \'echelle diagram
based on early {\it Kepler} data and, as noted by \citet{Montal2010},
has important diagnostic potential.
As discussed by Bedding (this volume)
this potential was dramatically realized in the analyses by \citet{Beck2011} and
\citet{Beddin2011}.
The groups of dipolar peaks near the acoustic resonances
were found to have a regular structure which allowed determination
of well-defined period spacings, reminiscent of the behaviour of g modes
(cf.\ Eq.~\ref{eq:gper}) and matching a corresponding behaviour in
the computed frequencies of stellar models.
Such behaviour has also been found in analysis of CoRoT data
\citep{Mosser2011b}.

\begin{figure}
\begin{center}
\includegraphics[width=7.5cm]{\fig/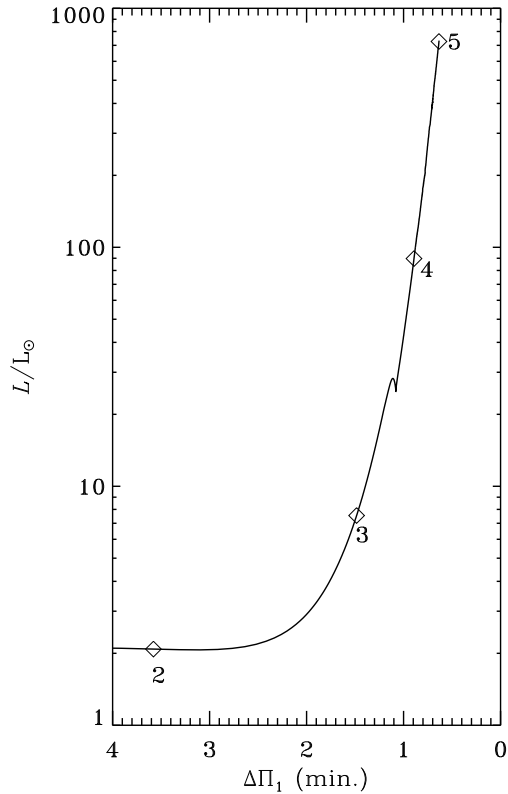}
\caption{`Hertzsprung-Russell diagram', in terms of the
asymptotic dipolar g-mode spacing $\Delta \Pi_1$ (cf.\ Eq.~\ref{eq:perspac1})
and luminosity, for the $1\,M_\odot$ evolution track illustrated in
Fig.~\ref{fig:hr1}.
}
   \label{fig:hrdpi1}
\end{center}
\end{figure}

To illustrate the properties of the computed period spacings
Fig.~\ref{fig:perspac} shows the difference in period between adjacent
dipolar modes in Model~3 in the $1 \, M_\odot$ evolution sequence
illustrated in Fig.~\ref{fig:hr1}.
It is clear that, particularly in the dense region of g-dominated modes
at relatively low frequency, there is an almost uniform spacing, 
interspersed with decreases which are associated with the acoustic resonances.
To illustrate this relation, the figure also shows $\log Q_{nl}$ where
$Q_{nl}$ is the mode inertia normalized by the inertia of a radial mode
at the same frequency.
Specifically,
\begin{equation}
Q_{nl} = {E_{nl} \over \bar E_0(\omega_{nl})} \; ,
\label{eq:qnl}
\end{equation}
where $E_{nl}$ is the inertia of the mode and $\bar E_0(\omega_{nl})$ is
the inertia of radial modes, interpolated to the frequency $\omega_{nl}$ 
of the mode.
It is obvious that the variation in the period spacing is closely related
to the variation in the inertia.

The uniform period spacing is clearly only realized for the most g-dominated
modes whose high inertia makes detection unlikely.%
\footnote{except perhaps in extremely long timeseries, exceeding 
the lifetimes of even these modes}
However, it was demonstrated by \citet{Beddin2011} that with the
{\it Kepler} data the pure g-mode period spacing can be determined
from the observed frequencies in data of sufficient quality;
this was illustrated in a `g-mode \'echelle diagram', dividing the
spectrum into segments of fixed length in period (see also Bedding,
this volume).
In other cases the period spacings can be inferred from analysis of 
the power spectra and extrapolated to the value for pure g modes
\citep[see also][]{Mosser2011b}.
\citet{Beddin2011} showed that the inferred period spacing is substantially
higher for clump stars in the core helium burning phase than for
stars on the ascending red-giant branch, providing a clear separation into
these two groups of stars that are superficially very similar.

\begin{figure}
\begin{center}
\includegraphics[width=8.0cm]{\fig/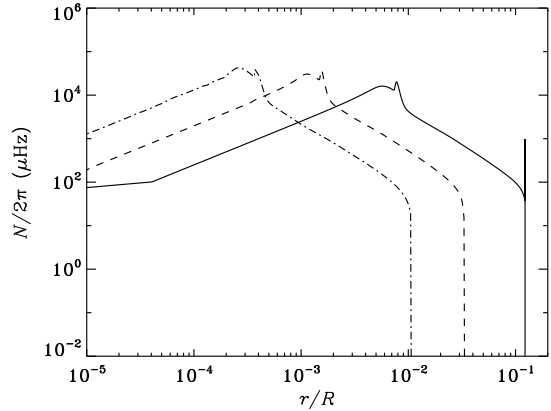}
\caption{Cyclic buoyancy frequencies $N/2 \pi$ in
Model~3 (solid), Model~4 (dashed) and Model~5 (dot-dashed)
in the $1\,M_\odot$ evolution sequence illustrated in Figs~\ref{fig:hr1}
and \ref{fig:hrdpi1}.
}
   \label{fig:bv1}
\end{center}
\end{figure}

{}From the asymptotic results in Eqs~(\ref{eq:gper}) and (\ref{eq:gpersp})
the period spacing for pure dipolar g modes is approximately given by
\begin{equation}
\Delta \Pi_1 = \sqrt{2} \pi^2 \left(\int_{r_1}^{r_2} N 
{\dd r \over r}\right)^{-1} \; .
\label{eq:perspac1}
\end{equation}
This is illustrated by the dashed horizontal line in Fig.~\ref{fig:perspac};
it is clearly in good agreement with the value obtained from the most
strongly trapped modes.
Given the $\Delta \Pi_1$, as discussed above, can be obtained
from the observations, 
we therefore obtain diagnostics of the buoyancy frequency in the core of the
star.

This can be used to illustrate the evolution of the star in an `HR' diagram,
plotting $L$ against $\Delta \Pi_1$, as done in Fig.~\ref{fig:hrdpi1}.
It is evident that $\Delta \Pi_1$ varies little as the star moves up the
red-giant branch.
This behaviour can be understood in terms of the variation in the 
buoyancy frequency, illustrated for three representative models in 
Fig.~\ref{fig:bv1}.
On this logarithmic $r$ scale the main effect is a shift towards smaller
radii as the core contracts, with little change in the shape of $N$.
Noting that the integral in Eq.~(\ref{eq:perspac1}) is in terms of $\log r$
it follows that there is little change in $\Delta \Pi_1$ with evolution.
It should also be noticed that $N$ is depressed in the core of the star.
This is probably caused by the increasing electron degeneracy of the gas,
leading to $\dd \ln p / \dd \ln \rho$ approaching $5/3$.%
\footnote{In white dwarfs this leads to a very small buoyancy frequency
in the interior of the star, confining the g modes to the outer layers
\citep{Fontai2008}.}

\begin{figure}
\begin{center}
\includegraphics[width=8.0cm]{\fig/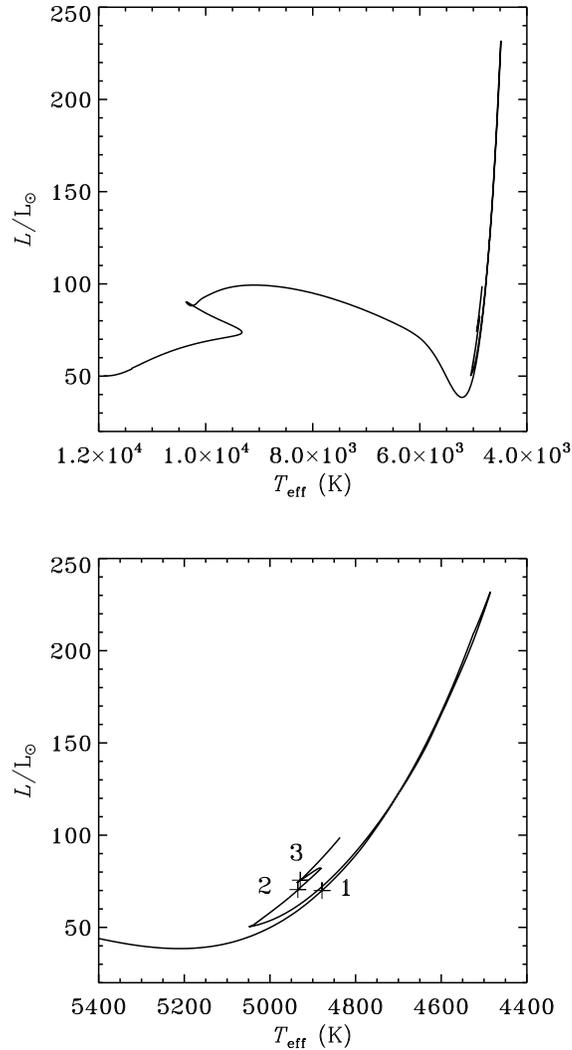}
\caption{Evolution track for a $2.5\,M_\odot$ model.
The track goes through helium ignition at the tip of the red-giant branch and 
ends near central helium exhaustion.
}
   \label{fig:hr25}
\end{center}
\end{figure}

As discussed by Bedding (this volume) there is a striking 
difference in $\Delta \Pi_1$ between stars in the hydrogen shell-burning
phase, as the $1 \, M_\odot$ model illustrated above, and models that burn
helium in the core.
Unfortunately ASTEC does not allow computation of a model through
a helium flash, and hence the evolution of the $1 \, M_\odot$ model cannot
be followed beyond helium ignition.
For more massive stars the helium ignition takes place in a more quiet manner
which can be followed by ASTEC. 
Figure~\ref{fig:hr25} shows the evolution track of a $2.5 \, M_\odot$ model
up to a point near exhaustion of helium at the centre.

\begin{figure}
\begin{center}
\includegraphics[width=8.0cm]{\fig/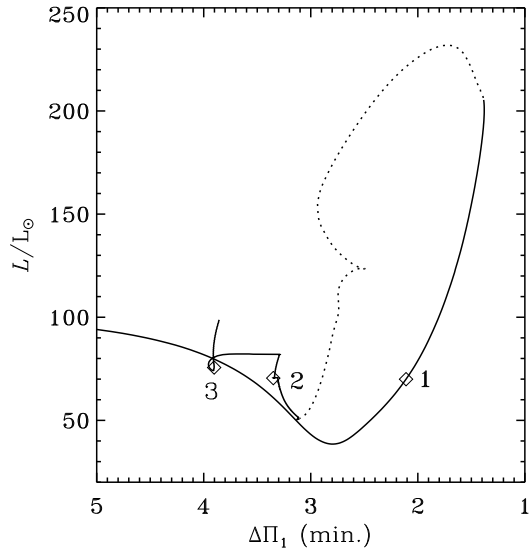}
\caption{`Hertzsprung-Russell diagram', in terms of the
asymptotic dipolar g-mode spacing $\Delta \Pi_1$ (cf.\ Eq.~\ref{eq:perspac1})
and luminosity, for the $2.5\, M_\odot$ evolution track illustrated in
Fig.~\ref{fig:hr25}.
The dotted part of the curve extends from helium ignition, at 
a model age of 0.421\,Gyr,
to the establishment of stable helium burning, at 0.439\,Gyr.
(The end of the track is at 0.565 Gyr.)
}
   \label{fig:hrdpi25}
\end{center}
\end{figure}

In the $(\Delta \Pi_1, L)$ diagram this leads to a rather more complex 
evolution, illustrated in Fig.~\ref{fig:hrdpi25}.
The variation on the ascending red-giant branch is very similar to the
$1 M_\odot$ evolution illustrated in Fig.~\ref{fig:hrdpi1}.
With the helium ignition, the star moves through a somewhat convoluted
path, shown dotted in the figure, towards larger $\Delta \Pi_1$,
before the star settles down to quiet helium burning.%
\footnote{The additional variations along this path appear to be related 
to modest oscillations in the central properties, similar to, but of smaller 
amplitude, than the oscillations seen in a full-fledged helium flash
\citep{Serene2005}.}
During the central helium burning $\Delta \Pi_1$ undergoes a further jump to 
smaller larger values, which is also reflected in the evolution track in
Fig.~\ref{fig:hr25}.

\begin{figure}
\begin{center}
\includegraphics[width=7.0cm]{\fig/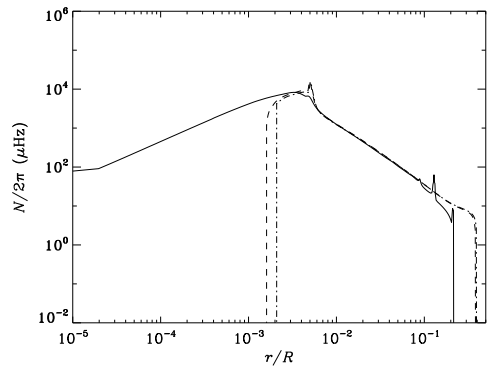}
\caption{Cyclic buoyancy frequencies $N/2 \pi$ in
Model~1 (solid), Model~2 (dashed) and Model~3 (dot-dashed)
in the $2.5\,M_\odot$ evolution sequence illustrated in Figs~\ref{fig:hr25}
and \ref{fig:hrdpi25}.
}
   \label{fig:bv25}
\end{center}
\end{figure}

Again, the variation in $\Delta \Pi_1$ can be understood by considering the
evolution of the buoyancy frequency, illustrated in Fig.~\ref{fig:bv25}.
Comparing Models 1 and 2, at roughly the same luminosity, the former model
is obviously similar to the red-giant $1 M_\odot$ model in Fig.~\ref{fig:bv1}.
In Model~2, on the other hand, helium burning causes a convective core
where the g modes are excluded; since the behaviour of $N$ in the rest of the
core is quite similar to that of Model~1, the effect is to decrease the integral
in Eq.~(\ref{eq:perspac1}) and hence to increase $\Delta \Pi_1$.
This is the dominant cause of the remarkable difference in the period
spacing between stars on the red-giant branch and in the red clump
observed by \citet{Beddin2011} and discussed further by Bedding in this volume.
The subsequent jump in $\Delta \Pi_1$ and $L$ is caused by a jump in
the size of the convective core, as illustrated 
in Fig.~\ref{fig:bv25} by Model~3,
probably associated with changes in the composition profile in the core
\citep[e.g.,][]{Castel1971}.

This analysis provides a simple explanation for the behaviour found 
by \citet{Beddin2011} and furthermore indicates how this may allow the
period spacing to be used as a sensitive diagnostics of the buoyancy frequency
in the stellar core.
Further investigations will be required to obtain a deeper understanding
of the diagnostic power of such data, including the ability to probe effects
of convective overshoot and rotationally induced mixing in the core.

\begin{acknowledgments}
I am very grateful to Pere Pall\'e and his colleagues at the IAC for the
organization of an excellent Winter School.
I look forward to the next opportunity to participate in such a school.
J. De Ridder and M.-A. Dupret are thanked for providing several key figures.
\end{acknowledgments}



\begin{thebibliography}{}




\bibitem[Aerts {et~al.}(2010)]{Aerts2010}
Aerts, C., Christensen-Dalsgaard, J. \& Kurtz, D. W. 2010,
{\it Asteroseismology}, Springer, Heidelberg

\bibitem[Angulo {et~al.}(1999)]{Angulo1999}
Angulo, C., Arnould, M., Rayet, M., Descouvemont, P., Baye, D., et al. 1999,
\titnt [A compilation of charged-particle induced thermonuclear reaction rates].
{\it Nucl. Phys. A}, {\rm 656}, 3 

\bibitem[Baglin {et~al.}(2009)]{Baglin2009}
Baglin, A., Auvergne, M., Barge, P., Deleuil, M., Michel, E. and
the CoRoT Exoplanet Science Team 2009,
\titnt [CoRoT: Description of the mission and early results].
in {\it Proc. IAU Symp. 253, Transiting Planets},
eds F. Pont, D. Sasselov \& M. Holman,
IAU and Cambridge University Press, p. 71 

\bibitem[Balmforth(1992)]{Balmfo1992}
Balmforth, N. J. 1992,
\titnt [Solar pulsational stability. I: Pulsation-mode thermodynamics].
{\it MNRAS}, {\rm 255}, 603 

\bibitem[Barban {et~al.}(2007)]{Barban2007}
Barban, C., Matthews, J. M., De Ridder, J., Baudin, F., Kuschnig, R., et al.
2007,
\titnt [Detection of solar-like oscillations in the red giant $\epsilon$ Ophiuchi
by MOST spacebased photometry].
{\it A\&A}, {\rm 468}, 1033 

\bibitem[Basu {et~al.}(2004)]{Basu2004}
Basu, S., Mazumdar, A., Antia, H. M. \& Demarque, P. 2004,
\titnt [Asteroseismic determination of helium abundance in stellar envelopes].
{\it MNRAS}, {\rm 350}, 277 

\bibitem[Basu {et~al.}(2011)]{Basu2011}
Basu, S., Grundahl, F., Stello, D., Kallinger, T., Hekker, S., et al. 2011,
\titnt [Sounding open clusters: asteroseismic constraints from {\it Kepler} on
the properties of NGC 6791 and NGC 6819].
{\it ApJ}, {\rm 729}, L10 

\bibitem[Batalha {et~al.}(2011)]{Batalh2011}
Batalha, N. M., Borucki, W. J., Bryson, S. T., Buchhave, L. A.,
Caldwell, D. A., et al. 2011,
\titnt [{\it Kepler}'s first rocky planet: Kepler-10b].
{\it ApJ}, {\rm 729}, 27 

\bibitem[Batchelor(1956)]{Batche1956}
Batchelor, G. K. 1956,
{\it The theory of homogeneous turbulence}.
Cambridge University Press

\bibitem[Baudin {et~al.}(2011)]{Baudin2011}
Baudin, F., Barban, C., Belkacem, K., Hekker, S., Morel, T., et al. 2011,
\titnt [Amplitudes and lifetimes of solar-like oscillations observed by CoRoT.
Red-giant versus main-sequence stars].
{\it A\&A}, {\rm 529}, A84 

\bibitem[Beck {et~al.}(2011)]{Beck2011}
Beck, P.~G., Bedding, T.~R., Mosser, B., Stello, D., Garcia, R.~A.,
et al. 2011,
\titnt [Kepler detected gravity-mode period spacings in a red giant].
{\it Science}, {\rm 332}, 205. 

\bibitem[Bedding \&  Kjeldsen(2003)]{Beddin2003}
Bedding, T. R. \& Kjeldsen, H. 2003,
\titnt [Solar-like oscillations].
{\it PASA}, {\rm 20}, 203 

\bibitem[Bedding {et~al.}(2010)]{Beddin2010}
Bedding, T.~R., Huber, D., Stello, D., Elsworth, Y.~P., Hekker, S.,
et al. 2010,
\titnt [Solar-like oscillations in low-luminosity red giants: first results from
{\it Kepler}].
{\it ApJ}, {\rm 713}, L176 

\bibitem[Bedding {et~al.}(2011)]{Beddin2011}
Bedding, T.~R., Mosser, B., Huber, D., Montalb{\'a}n, J., Beck, P.,
et al. 2011,
\titnt [Gravity modes as a way to distinguish between hydrogen- and helium-burning
red giant stars].
{\it Nature}, {\rm 471}, 608 

\bibitem[Belkacem {et~al.}(2006)]{Belkac2006}
Belkacem, K., Samadi, R., Goupil, M. J., Kupka, F. \& Baudin, F. 2006,
\titnt [A closure model with plumes. II. Application to the stochastic excitation
of solar $p$ modes].
{\it A\&A}, {\rm 460}, 183 

\bibitem[Belkacem {et~al.}(2011)]{Belkac2011}
Belkacem, K., Goupil, M. J., Dupret, M. A., Samadi, R., Baudin, F.,
Noels, A. \& Mosser, B. 2011,
\titnt [The underlying physical meaning of the $\nu_{\rm max} - \nu_{\rm c}$
relation].
{\it A\&A}, {\rm 530}, A142 

\bibitem[B{\"o}hm-Vitense(1958)]{Bohm1958}
B{\"o}hm-Vitense, E. 1958,
\titnt [\"Uber die Wasserstoffkonvektionszone in Sternen
verschiedener Effektivtemperaturen und Leuchtkr{\"a}fte].
{\it Z. Astrophys.}, {\rm 46}, 108 

\bibitem[Borucki {et~al.}(2009)]{Boruck2009}
Borucki, W., Koch, D., Batalha, N., Caldwell, D.,
Christensen-Dalsgaard, J., Cochran, W. D., Dunham, E., Gautier, T. N.,
Geary, J., Gilliland, R., Jenkins, J., Kjeldsen, H.,
Lissauer, J. J. \& Rowe, J. 2009,
\titnt [{\it KEPLER}: Search for Earth-size planets in the habitable zone].
in {\it Proc. IAU Symp. 253, Transiting Planets},
eds F. Pont, D. Sasselov \& M. Holman,
IAU and Cambridge University Press, p. 289 

\bibitem[Brown {et~al.}(1991)]{Brown1991}
Brown, T. M., Gilliland, R. L., Noyes, R. W. \& Ramsey, L. W. 1991,
\titnt [Detection of possible $p$-mode oscillations of Procyon].
{\it ApJ}, {\rm 368}, 599 

\bibitem[Castellani {et~al.}(1971)]{Castel1971}
Castellani, V., Giannone, P. \& Renzini, A. 1971,
\titnt [Overshooting of convective cores in helium-burning horizontal-branch stars. 
I].
{\it Ap\&SS}, {\rm 10}, 340 

\bibitem[Chang \&  Gough(1998)]{Chang1998}
Chang, H.-Y. \& Gough, D. O. 1998,
\titnt [On the power distribution of solar p modes].
{\it Solar Phys.}, {\rm 181}, 251 

\bibitem[Chaplin {et~al.}(1997)]{Chapli1997}
Chaplin, W. J., Elsworth, Y., Howe, R., Isaak, G. R., McLeod, C. P.,
Miller, B. A. \& New, R. 1997,
\titnt [The observation and simulation of stochastically excited solar p modes].
{\it MNRAS}, {\rm 287}, 51 

\bibitem[Chaplin {et~al.}(2002)]{Chapli2002}
Chaplin, W. J., Elsworth, Y., Isaak, G. R., Marchenkov, K. I., Miller, B. A.,
New, R., Pinter, B. \& Appourchaux, T. 2002,
\titnt [Peak finding at low signal-to-noise: low-$\ell$ solar acoustic 
eigenmodes at $n \le 9$ from the analysis of BiSON data].
{\it MNRAS}, {\rm 336}, 979 

\bibitem[Chaplin {et~al.}(2005)]{Chapli2005}
Chaplin, W. J., Houdek, G., Elsworth, Y., Gough, D. O., Isaak, G. R. \&
New, R. 2005,
\titnt [On model predictions of the power spectral density of radial solar p modes].
{\it MNRAS}, {\rm 360}, 859 

\bibitem[Chaplin {et~al.}(2011)]{Chapli2011}
Chaplin, W. J., Kjeldsen, H., Christensen-Dalsgaard, J., Basu, S., Miglio, A.,
et al. 2011,
\titnt [Ensemble asteroseismology of solar-type stars with the NASA Kepler mission].
{\it Science}, {\rm 332}, 213 

\bibitem[Christensen-Dalsgaard(2004)]{Christ2004}
Christensen-Dalsgaard, J. 2004,
\titnt [Physics of solar-like oscillations].
{\it Solar Phys.}, {\rm 220}, 137 

\bibitem[Christensen-Dalsgaard(2008a)]{Christ2008a}
Christensen-Dalsgaard, J. 2008a,
\titnt [ASTEC -- the Aarhus STellar Evolution Code].
{\it Ap\&SS},  {\rm 316}, 13 

\bibitem[Christensen-Dalsgaard(2008b)]{Christ2008b}
Christensen-Dalsgaard, J. 2008b,
\titnt [ADIPLS -- the Aarhus adiabatic pulsation package].
{\it Ap\&SS},  {\rm 316}, 113 

\bibitem[Christensen-Dalsgaard \&  Frandsen(1983)]{Christ1983}
Christensen-Dalsgaard, J. \& Frandsen, S. 1983,
\titnt [Stellar 5 min oscillations].
{\it Solar Phys.}, {\rm 82}, 469 

\bibitem[Christensen-Dalsgaard \&  Thompson(2011)]{Christ2011}
Christensen-Dalsgaard, J. \& Thompson, M. J. 2011,
\titnt [Stellar hydrodynamics caught in the act: Asteroseismology with CoRoT
and Kepler].
in {\it Proc. IAU Symposium 271: Astrophysical dynamics:
from stars to planets},
eds N. Brummell, A. S. Brun, M. S. Miesch \& Y. Ponty,
IAU and Cambridge University Press, in the press
{\tt [arXiv:1104.5191]}

\bibitem[Christensen-Dalsgaard {et~al.}(1989)]{Christ1989}
Christensen-Dalsgaard, J., Gough, D. O. \& Libbrecht, K. G. 1989,
\titnt [Seismology of solar oscillation line widths].
{\it ApJ}, {\rm 341}, L103 

\bibitem[Christensen-Dalsgaard {et~al.}(2001)]{Christ2001}
Christensen-Dalsgaard, J., Kjeldsen, H. \& Mattei, J. A. 2001,
\titnt [Solar-like oscillations of semiregular variables].
{\it ApJ}, {\rm 562}, L141 

\bibitem[Christensen-Dalsgaard {et~al.}(2010)]{Christ2010}
Christensen-Dalsgaard, J., Kjeldsen, H., Brown, T.~M., Gilliland, R.~L.,
Arentoft, T., Frandsen, S., Quirion, P.-O., Borucki, W.~J., Koch, D. \&
Jenkins, J.~M. 2010,
\titnt [Asteroseismic investigation of known planet hosts in the {\it Kepler} field].
{\it ApJ}, {\rm 713}, L164 

\bibitem[De Ridder {et~al.}(2006)]{DeRidd2006}
De Ridder, J., Barban, C., Carrier, F., Mazumdar, A., Eggenberger, P.,
Aerts, C., Deruyter, S. \& Vanautgaerden, J. 2006,
\titnt [Discovery of solar-like oscillations in the red giant $\epsilon$ Ophiuchi].
{\it A\&A}, {\rm 448}, 689 

\bibitem[De Ridder {et~al.}(2009)]{DeRidd2009}
De Ridder, J., Barban, C., Baudin, F., Carrier, F., Hatzes, A. P.,
et al. 2009,
\titnt [Non-radial oscillation modes with long lifetimes in giant stars].
{\it Nature}, {\rm 459}, 398 

\bibitem[Deubner \&  Gough(1984)]{Deubne1984}
Deubner, F.-L. \& Gough, D. O. 1984,
\titnt [Helioseismology: Oscillations as a diagnostic of the solar interior].
{\it ARAA}, {\rm 22}, 593 

\bibitem[Di Mauro {et~al.}(2011)]{DiMaur2011}
Di Mauro, M.~P., Cardini, D., Catanzaro, G., Ventura, R., Barban, C.,
et al. 2011,
\titnt [Solar-like oscillations from the depths of the red-giant star KIC 4351319
observed with {\it Kepler}].
{\it MNRAS}, in the press
{\tt [arXiv:1105.1076]}

\bibitem[Dupret {et~al.}(2009)]{Dupret2009}
Dupret, M.-A., Belkacem, K., Samadi, R., Montalban, J., Moreira, O.,
Miglio, A., Godart, M., Ventura, P., Ludwig, H.-G., Grigahc{\`e}ne, A.,
Goupil, M.-J., Noels, A. \& Caffau, E. 2009,
\titnt [Theoretical amplitudes and lifetimes of non-radial solar-like
oscillations in red giants].
{\it A\&A}, {\rm 506}, 57 

\bibitem[Duvall(1982)]{Duvall1982}
Duvall, T. L. 1982,
\titnt [A dispersion law for solar oscillations].
{\it Nature}, {\rm 300}, 242 

\bibitem[Dziembowski(1977)]{Dziemb1977}
Dziembowski, W. 1977,
\titnt [Oscillations of giants and supergiants].
{\it AcA}, {\rm 27}, 95 

\bibitem[Dziembowski \&  Soszy{\'n}ski(2010)]{Dziemb2010}
Dziembowski, W. A. \& Soszy{\'n}ski, I. 2010,
\titnt [Acoustic oscillations in stars near the tip of the red giant branch].
{\it A\&A}, {\rm 524}, A88 

\bibitem[Dziembowski {et~al.}(2001)]{Dziemb2001}
Dziembowski, W. A., Gough, D. O., Houdek, G. \& Sienkiewicz, R. 2001,
\titnt [Oscillations of $\alpha$ UMa and other red giants].
{\it MNRAS}, {\rm 328}, 601 

\bibitem[Edmonds \&  Gilliland(1996)]{Edmond1996}
Edmonds, P. D. \& Gilliland, R. L. 1996,
\titnt [K giants in 47 Tucanae: detection of a new class of variable stars].
{\it ApJ}, {\rm 464}, L157 

\bibitem[Fletcher {et~al.}(2006)]{Fletch2006}
Fletcher, S. T., Chaplin, W. J., Elsworth, Y., Schou, J. \& Buzasi, D. 2006,
\titnt [Frequency, splitting, linewidth and amplitude estimates of low-$\ell$ p modes
of $\alpha$ Cen A: analysis of Wide-Field Infrared Explorer photometry].
{\it MNRAS}, {\rm 371}, 935 

\bibitem[Fontaine \&  Brassard(2008)]{Fontai2008}
Fontaine, G. \& Brassard, P. 2008,
\titnt [The pulsating white dwarf stars].
{\it PASP}, {\rm 120}, 1043 

\bibitem[Frandsen {et~al.}(2002)]{Frands2002}
Frandsen, S., Carrier, F., Aerts, C., Stello, D., Maas, T., et al. 2002,
\titnt [Detection of solar-like oscillations in the G7 giant star $\xi$ Hya].
{\it A\&A}, {\rm 394}, L5 

\bibitem[Gai {et~al.}(2011)]{Gai2011}
Gai, N., Basu, S., Chaplin, W. J. \& Elsworth, Y. 2011,
\titnt [An in-depth study of grid-based asteroseismic analysis].
{\it ApJ}, {\rm 730}, 63 

\bibitem[Gilliland {et~al.}(2010)]{Gillil2010}
Gilliland, R. L., Brown, T. M., Christensen-Dalsgaard, J., Kjeldsen, H.,
Aerts, C., et al. 2010,
\titnt [{\it Kepler} asteroseismology program: Introduction and first results].
{\it PASP}, {\rm 122}, 131 

\bibitem[Goldreich \&  Keeley(1977)]{Goldre1977}
Goldreich, P. \& Keeley, D. A. 1977,
\titnt [Solar seismology. II. The stochastic excitation of the solar
$p$-modes by turbulent convection].
{\it ApJ}, {\rm 212}, 243 

\bibitem[Goldreich {et~al.}(1994)]{Goldre1994}
Goldreich, P., Murray, N. \& Kumar, P. 1994,
\titnt [Excitation of solar $p$-modes].
{\it ApJ}, {\rm 424}, 466 

\bibitem[Gough(1977)]{Gough1977}
Gough, D. O. 1977,
\titnt [Mixing-length theory for pulsating stars].
{\it ApJ}, {\rm 214}, 196 

\bibitem[Gough(1980)]{Gough1980}
Gough, D. O. 1980,
\titnt [Some theoretical remarks on solar oscillations].
in {\it Lecture Notes in Physics}, vol. {\rm 125},
eds H. A. Hill \& W. Dziembowski, Springer-Verlag, p. 273 

\bibitem[Gough(1986)]{Gough1986}
Gough, D. O. 1986,
\titnt [EBK quantization of stellar waves].
in {\it Hydrodynamic and magnetohydrodynamic problems in the Sun and stars},
ed. Y. Osaki, Department of Astronomy, University of Tokyo, p. 117 

\bibitem[Gough(1990)]{Gough1990}
Gough, D. O. 1990,
\titnt [Comments on helioseismic inference].
in {\it Progress of seismology of the sun and stars},
{\it Lecture Notes in Physics}, vol. {\rm 367}, 
eds Y. Osaki \& H. Shibahashi, Springer, Berlin, p. 283 

\bibitem[Gough(1993)]{Gough1993}
Gough, D. O. 1993,
\titnt [Course 7. Linear adiabatic stellar pulsation].
in {\it Astrophysical fluid dynamics, Les Houches Session XLVII},
eds J.-P. Zahn \& J. Zinn-Justin, Elsevier, Amsterdam, p. 399 

\bibitem[Gough(2007)]{Gough2007}
Gough, D. O. 2007,
\titnt [An elementary introduction to the JWKB theory].
{\it AN}, {\rm 328}, 273 

\bibitem[Grigahc{\`e}ne {et~al.}(2005)]{Grigah2005}
Grigahc{\`e}ne, A., Dupret, M.-A., Gabriel, M., Garrido, R. \& Scuflaire, R.,
2005.
\titnt [Convection-pulsation coupling. I. A mixing-length perturbative theory].
{\it A\&A}, {\rm 434}, 1055 

\bibitem[Hekker {et~al.}(2006)]{Hekker2006}
Hekker, S., Caerts, C., De Ridder, J. \& Carrier, F. 2006,
\titnt [Pulsations detected in the line profile variations of red giants. Modelling 
of line moments, line bisector and line shape].
{\it A\&A}, {\rm 458}, 931 

\bibitem[Hekker {et~al.}(2011)]{Hekker2011}
Hekker, S., Gilliland, R. L., Elsworth, Y., Chaplin, W. J., De Ridder, J.,
Stello, D., Kallinger, T., Ibrahim, K. A., Klaus, T. C. \& Li, J. 2011,
\titnt [Characterisation of red-giant stars in the public {\it Kepler} data].
{\it MNRAS}, in the press
{\tt [arXiv:1103.0141]}

\bibitem[Houdek(2010a)]{Houdek2010a}
Houdek, G. 2010a,
\titnt [Convection and oscillations].
{\it AN}, {\rm 331}, 998 

\bibitem[Houdek(2010b)]{Houdek2010b}
Houdek, G. 2010b,
\titnt [Stellar turbulence and mode physics].
{\it Ap\&SS}, {\rm 328}, 237 

\bibitem[Houdek \&  Gough(2002)]{Houdek2002}
Houdek, G. \& Gough, D. O. 2002,
\titnt [Modelling pulsation amplitudes of $\xi$ Hydrae].
{\it MNRAS}, {\rm 336}, L65 

\bibitem[Houdek \&  Gough(2007)]{Houdek2007}
Houdek, G. \& Gough, D. O. 2007,
\titnt [An asteroseismic signature of helium ionization].
{\it MNRAS}, {\rm 375}, 861 

\bibitem[Houdek {et~al.}(1999)]{Houdek1999}
Houdek, G., Balmforth, N. J., Christensen-Dalsgaard, J. \& Gough, D. O. 1999,
\titnt [Amplitudes of stochastically excited oscillations in main-sequence stars].
{\it A\&A}, {\rm 351}, 582 

\bibitem[Huber {et~al.}(2010)]{Huber2010}
Huber, D., Bedding, T. R., Stello, D., Mosser, B., Mathur, S.,
et al. 2010,
\titnt [Asteroseismology of red giants from the first four months of {\it Kepler}
data: global oscillation parameters for 800 stars].
{\it ApJ}, {\rm 723}, 1607 

\bibitem[Iglesias \&  Rogers(1996)]{Iglesi1996}
Iglesias, C. A. \& Rogers, F. J. 1996,
\titnt [Updated OPAL opacities].
{\it ApJ}, {\rm 464}, 943 

\bibitem[Innis {et~al.}(1988)]{Innis1988}
Innis, J. L., Isaak, G. R., Brazier, R. I., Belmonte, J. A., Palle, P. L.,
Roca Cortes, T. \& Jones, A. R. 1988,
\titnt [High precision velocity observations of Arcturus using
the 7699 {\AA} line of potassium].
in {\it Seismology of the Sun \& Sun-like Stars}, 
eds V. Domingo \& E. J. Rolfe, 
ESA SP-286, ESA Publications Division,
Noordwijk, The Netherlands, p. 569 

\bibitem[Kallinger {et~al.}(2010)]{Kallin2010}
Kallinger, T., Weiss, W. W., Barban, C., Baudin, F., Cameron, C.,
Carrier, F., De Ridder, J., Goupil, M.-J., Gruberbauer, M., Hatzes, A.,
Hekker, S., Samadi, R. \& Deleuil, M. 2010,
\titnt [Oscillating red giants in the CoRoT exofield: asteroseismic mass and
radius determination].
{\it A\&A}, {\rm 509}, A77 

\bibitem[Kippenhahn \&  Weigert(1990)]{Kippen1990}
Kippenhahn, R. \& Weigert, A. 1990,
{\it Stellar structure and evolution},
Springer-Verlag, Berlin

\bibitem[Kiss \&  Bedding(2003)]{Kiss2003}
Kiss, L. L. \& Bedding, T. R. 2003,
\titnt [Red variables in the OGLE-II data base -- I. Pulsations and period-luminosity
relations below the tip of the red giant branch of the Large Magellanic Cloud].
{\it MNRAS}, {\rm 343}, L79 

\bibitem[Kiss \&  Bedding(2004)]{Kiss2004}
Kiss, L. L. \& Bedding, T. R. 2004,
\titnt [Red variables in the OGLE-II data base -- II. Comparison of the Large and 
Small Magellanic Clouds].
{\it MNRAS}, {\rm 347}, L83 

\bibitem[Kjeldsen \&  Bedding(1995)]{Kjelds1995}
Kjeldsen, H. \& Bedding, T. R. 1995,
\titnt [Amplitudes of stellar oscillations: the implications for asteroseismology].
{\it A\&A}, {\rm 293}, 87 

\bibitem[Kjeldsen {et~al.}(2008)]{Kjelds2008}
Kjeldsen, H., Bedding, T. R. \& Christensen-Dalsgaard, J. 2008,
\titnt [Correcting stellar oscillation frequencies for near-surface effects].
{\it ApJ}, {\rm 683}, L175 

\bibitem[Kumar {et~al.}(1988)]{Kumar1988}
Kumar, P., Franklin, J. \& Goldreich, P. 1988,
\titnt [Distribution function for the time-averaged energies of
stochastically excited solar $p$-modes].
{\it ApJ}, {\rm 328}, 879 

\bibitem[Mazumdar(2005)]{Mazumd2005}
Mazumdar, A. 2005,
\titnt [Asteroseismic diagrams for solar-type stars].
{\it A\&A}, {\rm 441}, 1079 

\bibitem[Miglio {et~al.}(2009)]{Miglio2009}
Miglio, A., Montalb{\'a}n, J., Baudin, F., Eggenberger, P., Noels, A., 
Hekker, S., De Ridder, J., Weiss, W. \& Baglin, A. 2009,
\titnt [Probing populations of red giants in the galactic disk with CoRoT].
{\it A\&A}, {\rm 503}, L21 

\bibitem[Miglio {et~al.}(2010)]{Miglio2010}
Miglio, A., Montalb{\'a}n, J., Carrier, F., De Ridder, J., Mosser, B.,
Eggenberger, P., Scuflaire, R., Ventura, P., D'Antona, F., Noels, A. \&
Baglin, A. 2010,
\titnt [Evidence for a sharp structure variation inside a red-giant star].
{\it A\&A}, {\rm 520}, L6 

\bibitem[Montalb{\'a}n {et~al.}(2010)]{Montal2010}
Montalb{\'a}n, J., Miglio, A., Noels, A., Scuflaire, R. \& Ventura, P. 2010,
\titnt [Seismic diagnostics of red giants: first comparison with stellar models].
{\it ApJ}, {\rm 721}, L182 

\bibitem[Monteiro \&  Thompson(2005)]{Montei2005}
Monteiro, M. J. P. F. G. \& Thompson, M. J. 2005,
\titnt [Seismic analysis of the second ionization region of helium in the Sun -- I.
Sensitivity study and methodology].
{\it MNRAS}, {\rm 361}, 1187 

\bibitem[Mosser {et~al.}(2010)]{Mosser2010}
Mosser, B., Belkacem, K., Goupil, M.-J., Miglio, A., Morel, T., Barban, C.,
Baudin, F., Hekker, S., Samadi, R., De Ridder, J., Weiss, W., Auvergne, M. \&
Baglin, A. 2010,
\titnt [Red-giant seismic properties analyzed with CoRoT].
{\it A\&A}, {\rm 517}, A22 

\bibitem[Mosser {et~al.}(2011a)]{Mosser2011a}
Mosser, B., Belkacem, K., Goupil, M. J., Michel, E., Elsworth, Y., 
et al. 2011a,
\titnt [The universal red-giant oscillation pattern. An automated determination 
with CoRoT data].
{\it A\&A}, {\rm 525}, L9 

\bibitem[Mosser {et~al.}(2011b)]{Mosser2011b}
Mosser, B., Barban, C., Montalb{\'a}n, J., Beck, P. G., Miglio, A.,
et al. 2011b,
\titnt [Mixed modes in red-giant stars observed with CoRoT].
{\it A\&A}, in the press
{\tt [arXiv:1105.6113v2]}

\bibitem[Retter {et~al.}(2003)]{Retter2003}
Retter, A., Bedding, T. R., Buzasi, D. L., Kjeldsen, H. \& Kiss, L. L. 2003,
\titnt [Oscillations in Arcturus from {\it WIRE} photometry].
{\it ApJ}, {\rm 591}, L151 
(Erratum: {\it ApJ}, {\rm 596}, L125)

\bibitem[Rogers {et~al.}(1996)]{Rogers1996}
Rogers, F. J., Swenson, F. J. \& Iglesias, C. A. 1996,
\titnt [OPAL Equation-of-State Tables for Astrophysical Applications].
{\it ApJ}, {\rm 456}, 902 

\bibitem[Samadi {et~al.}(2007)]{Samadi2007}
Samadi, R., Georgobiani, D., Trampedach, R., Goupil, M. J., Stein, R. F. \&
Nordlund, {\AA}. 2007,
\titnt [Excitation of solar-like oscillations across the HR diagram].
{\it A\&A}, {\rm 463}, 297 

\bibitem[Serenelli \&  Weiss(2005)]{Serene2005}
Serenelli, A. \& Weiss, A. 2005,
\titnt [On constructing horizontal branch models].
{\it A\&A}, {\rm 442}, 1041 

\bibitem[Smith {et~al.}(1987)]{Smith1987}
Smith, P. H., McMillan, R. S. \& Merline, W. J. 1987,
\titnt [Evidence for periodic radial velocity variations in Arcturus].
{\it ApJ}, {\rm 317}, L79 

\bibitem[Soszy{\'n}ski {et~al.}(2007)]{Soszyn2007}
Soszy{\'n}ski, I., Dziembowski, W. A., Udalski, A., Kubiak, M.,
Szyma{\'n}ski, M. K., Pietrzy{\'n}ski, G., Wyrzykowski, {\L}., Szewczyk, O. \&
Ulaczyk, K. 2007,
\titnt [The Optical Gravitational Lensing Experiment. Period-Luminosity
relations of variable red giant stars].
{\it AcA}, {\rm 57}, 201 

\bibitem[Stein(1968)]{Stein1968}
Stein, R. F. 1968,
\titnt [Waves in the solar atmosphere. I. The acoustic energy flux].
{\it ApJ}, {\rm 154}, 297 

\bibitem[Stello {et~al.}(2008)]{Stello2008}
Stello, D., Bruntt, H., Preston, H. \& Buzasi, D. 2008,
\titnt [Oscillating K giants with the {\it WIRE\/} satellite: determination of their
asteroseismic masses].
{\it ApJ}, {\rm 674}, L53 

\bibitem[Tabur {et~al.}(2010)]{Tabur2010}
Tabur, V., Bedding, T. R., Kiss, L. L., Giles, T., Derekas, A. \& Moon, T. T.
2010,
\titnt [Period-luminosity relations of pulsating M giants 
in the solar neighbourhood and the Magellanic Clouds].
{\it MNRAS}, {\rm 409}, 777 

\bibitem[Tassoul(1980)]{Tassou1980}
Tassoul, M. 1980,
\titnt [Asymptotic approximations for stellar nonradial pulsations].
{\it ApJS}, {\rm 43}, 469 

\bibitem[Tassoul(1990)]{Tassou1990}
Tassoul, M. 1990,
\titnt [Second-order asymptotic approximations for stellar nonradial acoustic modes].
{\it ApJ}, {\rm 358}, 313 

\bibitem[Teixeira {et~al.}(2003)]{Teixei2003}
Teixeira, T. C., Christensen-Dalsgaard, J., Carrier, F., Aerts, C.,
Frandsen, S., et al. 2003,
\titnt [Giant vibrations in dip].
{\it Ap\&SS}, {\rm 284}, 233 

\bibitem[Unno(1967)]{Unno1967}
Unno, W. 1967,
\titnt [The stellar radial pulsation coupled with the convection].
{\it PASJ}, {\rm 19}, 140 

\bibitem[Vandakurov(1967)]{Vandak1967}
Vandakurov, Yu. V. 1967,
\titnt [The frequency distribution of stellar oscillations].
{\it Astron. Zh.}, {\rm 44}, 786 
(English translation: {\it Soviet Astronomy AJ}, {\rm 11}, 630) 

\bibitem[Vorontsov {et~al.}(1991)]{Voront1991}
Vorontsov, S. V., Baturin, V. A. \& Pamyatnykh, A. A. 1991,
\titnt [Seismological measurement of solar helium abundance].
{\it Nature}, {\rm 349}, 49 

\bibitem[Xiong {et~al.}(1997)]{Xiong1997}
Xiong, D. R., Chen, Q. L. \& Deng, L. 1997,
\titnt [Nonlocal time-dependent convection theory].
{\it ApJS}, {\rm 108}, 529 

\bibitem[Xiong \&  Deng(2010)]{Xiong2010}
Xiong, D. R. \& Deng, L. 2010,
\titnt [Non-adiabatic oscillations of the low- and intermediate-degree modes of the
Sun].
{\it MNRAS}, {\rm 405}, 2759 

\end{thebibliography}
\end{document}